\documentclass{IEEEtran}
\usepackage{ifpdf}
\usepackage{cite}
\usepackage{amsmath}
\usepackage{algorithmic}
\usepackage{array}
\usepackage{stfloats}
\usepackage{url}
\usepackage{color}
\usepackage{multirow}

\ifCLASSINFOpdf
\usepackage[pdftex]{graphicx}
\else
\usepackage[dvips]{graphicx}
\fi

\begin{document}

%\title{\begin{Huge}High Performance Simultaneous Information and Power RB System by Telescope and Semiconductor\end{Huge}}
\title{\begin{Huge}High-Efficiency Resonant Beam Charging and Communication\end{Huge}}
\author{
Yunfeng~Bai,~%~\IEEEmembership{Student Member,~IEEE,} 
Qingwen~Liu,~\IEEEmembership{Senior Member,~IEEE,}
Xin~Wang,~\IEEEmembership{Fellow,~IEEE,}\\
Yudan Gou,
Bin Zhou and Zhiyong Bu
%Jun~Wu,~\IEEEmembership{Senior Member,~IEEE,}
%\IEEEauthorrefmark{1}
	
\thanks{Y.~Bai, and Q.~Liu, are with the College of Electronics and Information Engineering, Tongji University, Shanghai, China, (email: baiyf@tongji.edu.cn, qliu@tongji.edu.cn).}%
\thanks{Xin Wang is with the Key Laboratory for Information Science of Electromagnetic Waves, Department of Communication Science and Engineering, Fudan University, Shanghai 200433, China (e-mail: xwang11@fudan.edu.cn).}
\thanks{Yudan Gou is with the College of Electronics and Information Engineering, Sichuan University, Chengdu 610065, China(gouyudan@scu.edu.cn).}
\thanks{B.~Zhou, and Z.~Bu, are with the Key Laboratory of Wireless Sensor Network and Communications, Shanghai Institute of Microsystem and Information Technology, Chinese Academy of Sciences, Shanghai, China, (email: bin.zhou@mail.sim.ac.cn, zhiyong.bu@mail.sim.ac.cn).}
}

\maketitle

\begin{abstract}
%Simultaneous Wireless Information and Power Transfer (SWIPT) has recently been envisioned as an enabling technology capable of solving the power supply and data rate on the Internet of Things (IoT). In this paper, we proposed a high-efficiency resonant beam charging and communication system based on the telescope internal modulator (TIM) and semiconductor gain. To characterize the structure and output performance of the system, we establish the beam transfer, power cycle, and data reception models of the system. By systematically analyzing both system structure, energy harvester, and information receiver, we evaluate the system’s cavity stability, beam spot, energy output, and spectral efficiency. Numerical results illustrate that the system is stable which can realize long-range digital, and energy simultaneous transmission. Compared with the original system, the system has a better energy output capacity. Its threshold is lower and conversion efficiency is increased by 4 times. The system's spectrum efficiency can reach 45bit/s/Hz, which proves a good communication capability. %Overall, the system proposed provides a feasible long-range SWIPT scheme for IoT.
With the development of Internet of Things (IoT), demands of power and data for IoT devices increase drastically. In order to resolve the supply-demand contradiction, simultaneous wireless information and power transfer (SWIPT) has been envisioned as an enabling technology by providing high-power energy transfer and high-rate data delivering concurrently. 
In this paper, we introduce a high-efficiency resonant beam (RB) charging and communication scheme. The scheme utilizes the semiconductor materials as gain medium, 
%and the telescope internal modulator (TIM). 
which has a better energy absorption capacity compared with the traditional solid-state one. 
%Moreover, to match the gain size and reduce the transmission loss, the telescope internal modulator (TIM) are adopted in the scheme, which can \textcolor{blue}{efficiently focus the beam into the chip. }
Moreover, \textcolor{blue}{the telescope internal modulator (TIM) are adopted in the scheme which can concentrate beams to match the gain size, reducing the transmission loss.} 
%concentrate beams. 
%into a small-size gain medium, thus the long-distance separation cavity for SWIPT can be achieved. 
%As a result, the overall energy conversion efficiency can be enhanced over long-range beam transmission.
To evaluate the scheme SWIPT performance, we establish an analytical model and study the influence factors of its beam transmission, energy conversion, output power, and spectral efficiency. 
Numerical results shows that the proposed RB system can realize 16 W electric power output with 11 $\text{\%}$ end-to-end conversion efficiency, and support 18 bit/s/Hz spectral efficiency for communication. 

%stable SWIPT over 10 meters, whose energy conversion efficiency is increased by 13 times compared with the original system using the solid-state gain medium without TIM, and the spectrum efficiency can be above 18 bit/s/Hz. 

\end{abstract}


\IEEEpeerreviewmaketitle
%%%%%%%%%%%%%%%%%%%%%%%%%%%%%%%%%%%%%%%%%%%%%%%%%%%%%%%%%%%%%%%%%%%%%%%%%%%%%%%%%%%%%%%%%%%%%%%%%%%%%%%%%%%%%%%%%%%%
\section{Introduction}\label{one}
%In the era of 5G, with the gradual increase in application requirements, network capacity will continue to increase and provide services for at least 50 billion devices through wireless communication \cite{series2015imt}. At the same time, thanks to the rapid development of 5G, the internet of thing (IoT) industry is also prospering day by day. Its application scenarios become more abundant and more complex, which brings larger computing requirements. Energy consumption has increased sharply \cite{jin2018wireless}. Facing the aforementioned demands and challenges, various promising technologies are promoted. In the communication field, millimeter-wave transmission, ultra-dense cloud radio access networks, and massive multiple-input multiple-output (M-MIMO) arrays developed rapidly \cite{pan2018user,pan2017joint}. In terms of energy consumption, in addition to energy-efficient computing technology that reduces the amount of calculation per unit, wireless power transfer (WPT) is undoubtedly one of the most concerned technologies \cite{hui2013critical,lu2015wireless}. WPT is a technology that transmits electrical energy from the transmitter to the receiver of the electrical load without any physical connection. Compared with wired functions, WPT is more flexible which is suitable for mobile portable devices; compared with battery power, it is not limited by battery capacity, mass, volume, and battery replacement \cite{huang2017waveform,scrosati2010lithium}. 

Accompany by the growth of the Internet of Things (IoT), network capacity continues to increase due to rapidly growing number of devices \cite{series2015imt}. At the same time, the energy consumption of mobile devices is also increasing dramatically to support high-performance communication and computation \cite{jin2018wireless}. 
Facing these challenges, various technologies have been promoted, e.g., wireless power transfer (WPT), ultra-dense cloud radio access networks, etc. \cite{hui2013critical,lu2015wireless}.  Among them, WPT for providing unlimited energy supply is undoubtedly one of the most attractive solutions \cite{Lim19}. Compared with wired power supply, WPT is flexible and suitable for mobile devices; compared with battery, WPT is not restricted by battery’s capacity, weight, or volume \cite{huang2017waveform,scrosati2010lithium}.

%On this background, simultaneous wireless information and power transfer (SWIPT) technology has recently attracted wide attention from researchers as a means to provide both the information and energy for equipment terminals at the same time. Typical SWIPT technology is mainly divided into two types: wide-area omnidirectional and narrow-area orientation. Wide-area omnidirectional technology such as radio-wave \cite{shinohara2014wireless} can support long-distance and omnidirectional SWIPT. However, the broad-spectrum energy emission results in a lack of energy concentration, making it difficult to achieve high-power energy transmission. Narrow-area orientation technology such as laser \cite{haken1970laser} can support high power transmission. But using the narrow electromagnetic beam always accompanies the challenge of mobile receiver positioning. %safety?

\begin{figure}[t]
	\centering
	\includegraphics[scale=0.35]{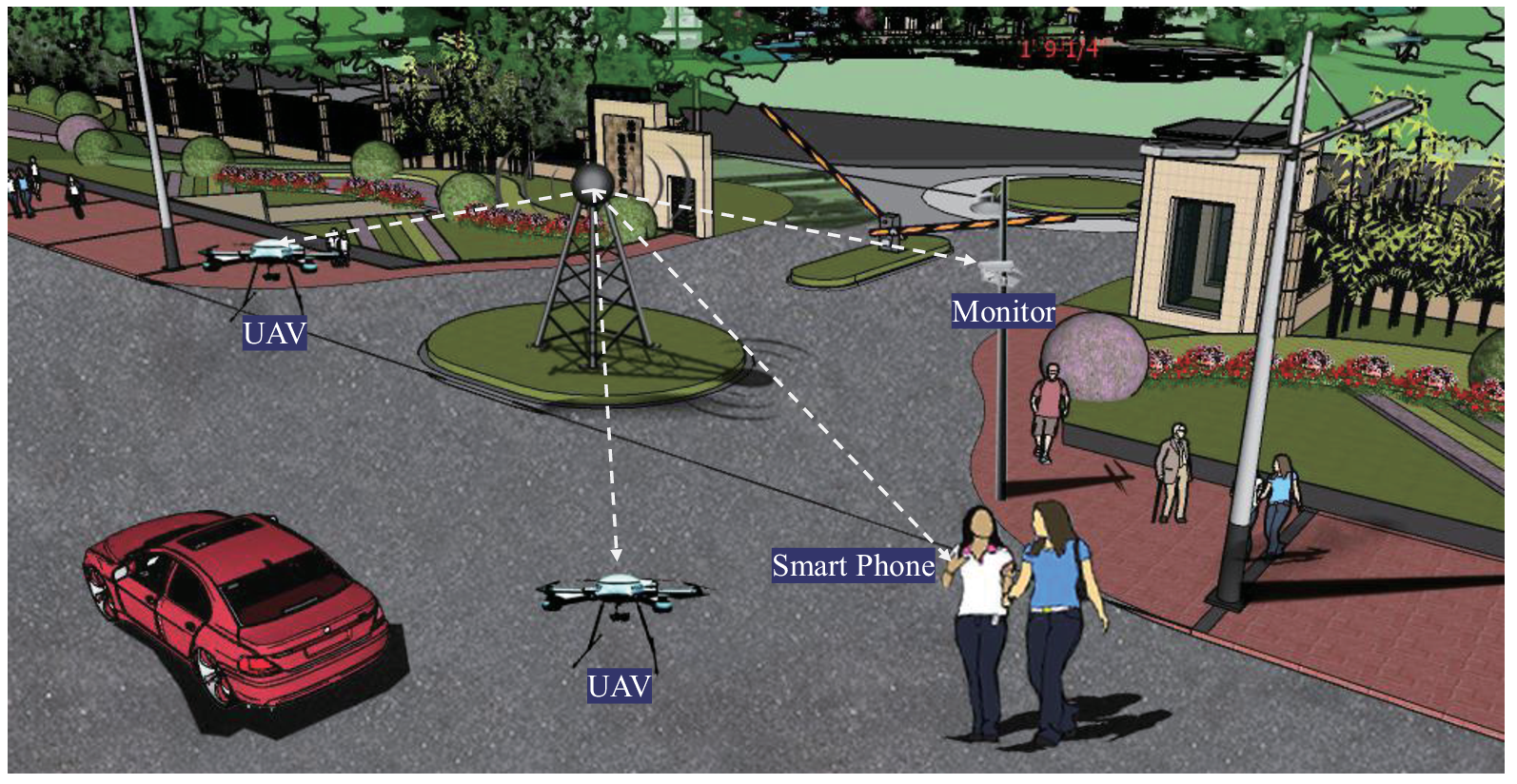}
	\caption{Application scenarios of SWIPT (UAV: unmanned aerial vehicle)}
	\label{Application}
\end{figure}

Based on the WPT, simultaneous wireless information and power transfer (SWIPT) technology has attracted great attention by researchers to address both the energy and data demands at the same time (Fig.~\ref{Application}). 
%SWIPT technology is divided into two types: wide-beam broadcast and narrow-beam orientation. Wide-beam broadcast such as radio frequency (RF) can support wide-coverage \cite{shinohara2014wireless}. 
%However, it faces the difficulty of achieving high-power transmission due to energy dissipation and radiation safety~\cite{habash2009recent}. Narrow-beam orientation technology such as light beam-forming can support high-power transmission. However, the narrow beam has the challenge of mobile position~\cite{haken1970laser}. 
SWIPT technologies have many forms with different system structure depending on the carrier. 
SWIPT system utilizing radio frequency (RF) as carrier is one of typical schemes. It depends on the mature RF communication technology, which has \textcolor{blue}{the characteristics of} low cost, long transfer distance, and expansive coverage. However, \textcolor{blue}{humans may be physical damage in the presence of high density RF radiation. Thus}, the researchers must gain more thorough understanding of potential health and safety impacts associated with the usage of SWIPT in public settings~\cite{habash2009recent}. 
Optical beams such as visible light \textcolor{blue}{can also be applied as the SWIPT carrier. The optical SWIPT scheme effectively harnesses the unique properties of light to support high-power energy transfer. This approach also offers significant communication advantages, including access to an extensive license-free spectrum, zero electromagnetic interference, and the capability for high-speed data transmission}
%Benefiting from the characteristic feature of the light, the optical SWIPT scheme usually has the potential to} support high power energy transfer, and has the great communication potential of huge license free spectrum, no electromagnetic interference, and high data rates 
\cite{elgala2011indoor,2016SPIE10158E}. %GPT润色
However, typical optical beams schemes are limited by alignment, safety, etc.
%due to the light divergence, its available coverage area of is limited.
%However, it faces the challenge of achieving high-power transmission due to energy dissipation and radiation safety~

As one of the optical beams scheme, resonant beam system (RBS) was proposed for self-alignment, and radiation safety \textcolor{blue}{light transmission} ~\cite{9653949,liu2016charging}.
%To meet the requirements of both high-power and high data rates, resonant beam system (RBS) was proposed \cite{zhang2018distributed}. 
The transmitter of the RBS is combined with the receiver to form a spatially separated resonator (SSR). The resonant beam generates in the cavity and carrying energy and information transfers between the transmitter and the receiver through the reflectors. Due to the beam concentration characteristics, RBS can support high power transfer~\cite{wang2019wireless}. Moreover, by cooperating retro-reflectors, RBS enables self-alignment between the transmitter and the receiver for mobile SWIPT~\cite{9425612}. When the light path between the transmitter and the receiver is blocked thanks to the separate resonant cavity structure, the cyclic reciprocating process of the resonant beam in the cavity will interrupt immediately, which prevents foreign objects from continuous radiative exposure, ensuring the human safety~\cite{fang2021safety}.
Besides, as the aforementioned statement, since the light beam is the signal carrier in the RBS, it can allow high-rate data transfer \cite{khalighi2014survey}. 
%In addition, the light beam has no interference with radio frequency and thus the RBS is environmental-friendly \cite{2016SPIE10158E}. 
Due to these remarkable advantages, the RBS has been applied in various SWIPT applications such as unmanned aerial vehicles (UAV), smartphones, \textcolor{blue}{monitors}, etc. \cite{chen2019resonant,liu2019trajectory}. 

The theoretical model, system design, deployment of the RBS have been studied in the literature recently. 
%Zhang \emph{et al.} in \cite{zhang2018distributed} introduced the beam transfer function and establish the analysis model of the RBS (a.k.a. distributed laser). 
%In addition to the theoretical analysis, 
Firstly, Zhang \emph{et al.} verified that the RBS can support 2 W WPT with 1 $\text{\%}$ end-to-end conversion efficiency in experiment \cite{wang2019wireless}. 
Sheng \emph{et al.} demonstrated an efficient, long-distance scheme using cat-eye reflectors to realize transmitter alignment, a telescope to concentrate beams, and adopting aspheric lenses to fix spherical aberration~\cite{Sheng21}. In \cite{zhang20222m}, Zhang \emph{et al.} characterized a long vertical external-cavity surface-emitting lasers resonator scheme for wireless charging. Moreover, to support multi-devices access into the RBS, Xiong \emph{et al.} proposed a time-division multiple access (TDMA) RBS design for wireless power transfer \cite{xiong2018tdma}. Xiong \emph{et al.} proposed the RB communication design and the analytical model in \cite{xiong2020resonant}. Liu \emph{et al.} presented the basic SWIPT model of the RBS and demonstrated its mobile ability using the retro-reflectors~\cite{9425612}. To evaluate the safety of the RBS, Fang \emph{et al.} proposed an analytical model based on electromagnetic field analysis and assessed its safety with external object invasion~\cite{fang2021safety}.

Previous research has underscored the \textcolor{blue}{features} and potential of Resonant Beam (RB) technology in Simultaneous Wireless Information and Power Transfer. Nevertheless, the original RB scheme still \textcolor{blue}{has} been challenged by significant energy conversion losses, resulting in a merely approximate 1$\%$ energy conversion efficiency across short distances \cite{wang2019wireless}. 
\textcolor{blue}{
The relatively low conversion efficiency of this technology presents challenges for its application in power supply scenarios for devices under a low-carbon and energy-saving context. Moreover, due to the suboptimal transmission efficiency, the system requires a greater input to satisfy the power demand at the output end. This situation imposes further challenges on system cooling, safety, and component durability.}
%低的转换损耗，使其不利于应用在低碳节能背景下的设备供能场景；同时低的传输损耗，使系统在满足输出端功率需求时，需提供更大的输入需求，这对于系统的散热，安全，元件的耐久也产生了挑战。

%within the gain medium and beam transmission losses

\textcolor{blue}{In this study, we present a novel RB-SWIPT scheme designed to elevate both power transmission efficiency and communication efficacy. We propose the use of a semiconductor gain medium, in contrast to the conventional solid-state medium. Semiconductors exhibit superior energy absorption capacity from the pump source, thus promoting a more effective energy level transition and population inversion efficiency. Consequently, the power threshold decreases, thereby improving the overall conversion efficiency.}

\textcolor{blue}{Additionally, we integrate a telescope internal modulator (TIM) into the resonator. The TIM serves to compress the beam, thereby harmonizing with the smaller dimensions of the gain medium and subsequently reducing transmission loss. This strategic modification underlines our commitment to enhancing the efficiency of SWIPT systems.}

%The semiconductor gain medium has a better energy absorption capacity compared with the traditional solid-state one, the overall energy conversion efficiency can be enhanced, while TIM can concentrate the diverged beam into a small-size gain medium, thus the long-distance separation cavity for SWIPT can be achieved. In order to enhance energy conversion efficiency, the semiconductor gain medium is introduced to replace the solid-state one. TIM can concentrate the diverged resonant beam into the small-sized gain medium, thus reduce the transmission loss, and increase the transmission efficiency. 

%Specifically, the telescope structure can modulate the light spot so that the light spot diverged due to the long-distance transmission can more effectively enter the small-sized gain module, which improves the optical feedback receiving efficiency, reduces the optical transmission loss, and increases the transmission distance.
%In order to solve the problem of low conversion efficiency, the semiconductor is introduced to replace the solid medium used in the present RB system. The semiconductor structural characteristics enable the gain to better absorb the energy of the pump source and the energy level transition and population inversion efficiency of the gain have also been enhanced which reduce the overall threshold of the system and prompt the conversion efficiency. 

The contributions of this paper are summarized as follows:
\begin{itemize}
    \item A high-efficiency RB-SWIPT scheme is proposed. By adopting the semiconductor gain medium and the telescope internal module, the scheme can achieve \textcolor{blue}{long-range}, high-rate, and \textcolor{blue}{energy} conversion efficiency enhanced SWIPT. %Specifically, we designed a telescope internal module (TIM) and placed it on the optical path in front of the gain module. When a light beam enters the telescope structure, the parallel lens changes the beam's longitudinal phase realizing beam compression, which leads to the transmission loss decrease. The gain module of the system is composed of semiconductor materials. When the pump source enters, most of the energy can be absorbed by the semiconductor medium and convert into photon lasing, which leads to the conversion loss decrease. 
    \item An analytical model of the proposed scheme is \textcolor{blue}{developed}, which can describe the energy conversion, the beam propagation, the electric power output, and data transfer capability \textcolor{blue}{of the system}. An evaluation for system performance and a guidance for parameters optimization is provided.
    %Specifically, we describe the propagation of optics with a matrix. When light passes through the air medium, lens, and other components, it is equivalent to a matrix change of the light vector. At this time, the stability of light propagation in the cavity and the changed spot size can be defined. What's more, we have established an energy output company based on semiconductor media, which can determine the parameters that affect energy conversion. We have also established a channel model of the system. %The mechanism of distance-enhanced and anti-interference ability are revealed. Mathematical models for the resonant cavity, energy conversion, and data transmission of the proposed scheme are established, which provides a feasible evaluation method for system performance and a guidance for parameters optimization.
    %\item We evaluated the energy and data transfer performance relying on the proposed analytical model, which demonstrates that the energy conversion efficiency gains 13 times and the spectrum efficiency can be above 20bit/s/Hz.  %The impacts of parameters such as the magnification of the TIM, the transmittance of the output mirror, and the splitting ratio on the simultaneous transmission performance are analyzed. The numerical results illustrate the feasibility of the proposed scheme and the advantages of its SWIPT characteristics. %These results offer the design guidance for practical engineering implementation for the RB system. 
\end{itemize}

%The rest of the paper is organized as follows. In Section II, the model of the high-efficiency RBS involving the semiconductor gain medium and TIM is described. In section III, the numerical evaluation of the system performance is depicted. The concerns on beam splitting process is discussed in Section IV. In section V, the conclusion is presented.
\textcolor{blue}{The remainder of this paper is structured as follows: Section II details the high-efficiency RBS model, including descriptions of the semiconductor gain medium and TIM. Section III presents a numerical evaluation of the system's performance. Concerns related to the beam splitting process are discussed in Section IV. Finally, Section V concludes the paper.}
%It involves the transmission distance, the beam, and the channel capacity which can give the transmission, energy, and communication capabilities of the system. 
%\section{System Fundamental Principle}

%\section{SYSTEM MODELS}\label{two}
%\subsection{System Structure}
\begin{figure}[t]
	\centering
	\includegraphics[scale=0.41]{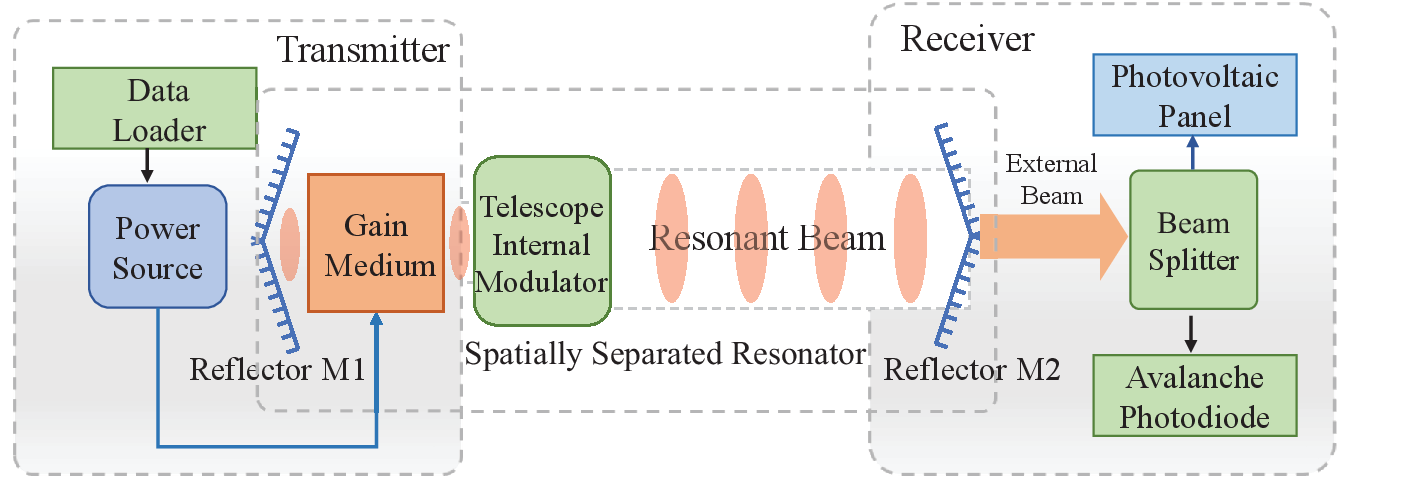}
	\caption{High-efficiency resonant beam SWIPT system
		%High efficiency design of resonant beam energy transmission system using telescope and semiconductor
	}
	\label{sys structure}
\end{figure}

\section{Scheme analysis and model establishment}
In this section, we will state the function principle of the scheme on realizing high-efficiency resonant beam charging and communication and develop the analytical model. 
\subsection{System Structure and Semiconductor Gain}
The system structure \textcolor{blue}{of the proposed RB-SWIPT scheme} is briefly depicted in Fig.~\ref{sys structure}. 
Overall, the system can be divided into three parts: transmitter, receiver, and spatially separated resonator. 
The transmitter is mainly composed of a gain medium, a pump source and a data loader. 
The data loader can load the electric signal into the pump source for data transfer. Then, the pump source releases beam carrying the data signal in the form of optical radiation to the gain medium. The gain medium receives the energy and occurs the stimulated absorption, spontaneous emission, and stimulated emission, generating photons to form resonant beams. 
%the optical radiation will function on the gain module where occur stimulated absorption, spontaneous emission and stimulated emission.
%The heat sink is used to absorb the heat generated on the gain module to suppress the influence of thermal effects on conversion efficiency. 
The receiver mainly includes three elements: photovoltaic (PV) panel module, beam splitter, and avalanche photodiode (APD) module \cite{aziz2014simulation,campbell2007APD}. The beam splitter \textcolor{blue}{is applied to} accept the external beams output from the M2 and split them into two rays. for \textcolor{blue}{independent} energy harvesting and data receiving. 
The PV module \textcolor{blue}{receives one of the rays for energy harvesting where the optical energy can be converted into electrical power after photoelectric conversion. The APD module captures the remaining rays for data reception, where the light signal is detected.} 
The spatially separated resonator is made by reflectors M1, M2, and a TIM. Among them, M1 and M2 constitute the resonant cavity for beams oscillation. 
%Subject to material properties, the size of the gain medium is usually relatively small in order to achieve high conversion efficiency. Thus, the beam size is usually greater than the gain medium, which causes unignorable beam loss. 
The TIM is placed on the optical path in front of the gain, which is used to change the phase of the resonant beam for beam compression and promoting the beam incidence.
\begin{figure}[t]
	\centering
	\includegraphics[scale=0.54]{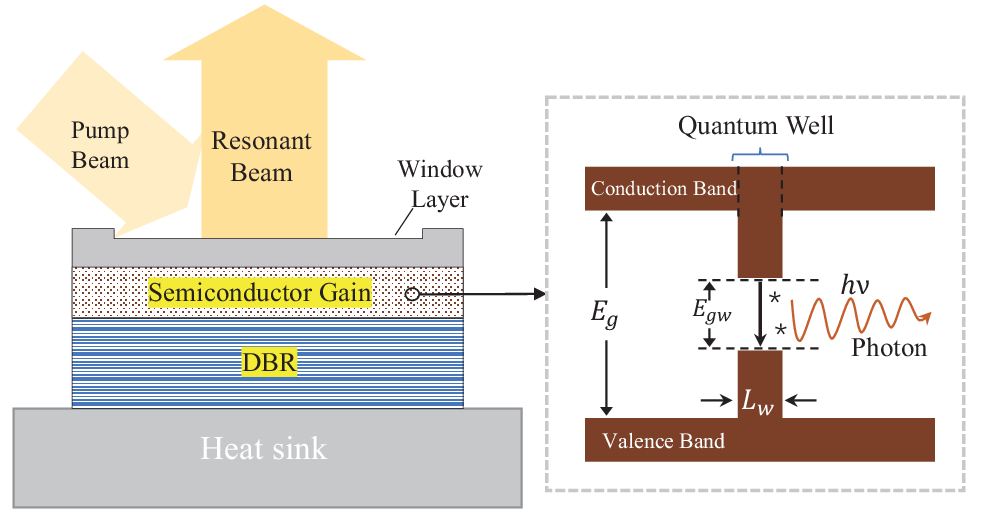}
	\caption{Transmitter structure and operating principles of the semiconductor gain (DBR: distributed Bragg reflector; $E_g$, $E_{gw}$: energy level difference; $L_w$: quantum well length; $hv$: photon energy)}
	\label{quantumwell}
\end{figure}

The proposed RB scheme applies the semiconductor materials to compose the gain medium for efficiency enhanced. 
Semiconductor material utilizes the recombination of electrons in the conduction band and holes in the valence band to generate simulated radiation~\cite{chow2012semiconductor}, 
%The semiconductor gain medium introduced herein has a hetero-junction quantum well structure. 
%Compared with the solid gain medium, 
which has the following characteristics: 
1) The materials with different ingredients on gain medium developing window layer, active area, and distributed Bragg reflector layer (DBR) have a band-gap difference forming a potential barrier. It will confine electrons and holes within the gain area, which is conducive to the recombination of electrons and holes; 
2) The active area and the cladding layers have a refractive index difference. When photons transfer on the material boundary, they will be reflected back to the active area. 
As a result, the light field can be restricted in the active area, which causes more stimulated radiation generation; 
3) The semiconductor gain can be made as a hetero-junction quantum well structure. This structure (Fig.~\ref{quantumwell}) can further reduce the thickness of the active layer to the nano-level producing the quantum effect, which makes the limitation of carriers and light fields be further strengthened, and causes the energy level difference decreasing from $E_g$ to $E_{gw}$ \cite{Soda_1979,soda1983gainasp}. 
%Overall, thanks to these characteristics, the loss of the particle conversion in semiconductor gain can be quite small, and the threshold value will be greatly reduced.
\textcolor{blue}{Thanks to these characteristics, the particle conversion loss in the semiconductor gain is significantly minimized, resulting in a substantial reduction of the threshold value.}
 %Note that no matter one-to-many scenario or one-to-one scenario, the basal principle of transmission process is same. Thus, a one-to-one system model and performance analysis will only conduct in this paper.
%thereby promoting the incidence of the beam and suppressing the overflow of the beam.
%more beam can be injected and seldom beam will overflow. %Promote the incidence of light and suppress the overflow of light
%The device includes a transmitter and a receiver. The transmitter is provided with a first reflector (M1) and a gain module. A second reflector (M2) and photoelectric conversion cells are sequentially arranged in the receiver, a resonant optical resonator is formed between the first reflector and the second reflector. The modulator, the first retroreflector, the gain module, and the telescope optical modulator are arranged in sequence along the optical path. information reception,detection.
%The telescope internal modulator includes a concave lens (L1) and a convex lens (L2) that are placed in parallel and the center of the lens is collinear. The focus of the convex lens in the direction of the concave lens coincides with the negative focus of the concave lens in the same direction. When the incident resonance beam passes through the convex lens, it overlaps. The focal point of the lens is converged, and after passing through the concave lens, a compressed beam of parallel output is formed to the gain module.

\subsection{Resonant Beam Power and Conversion Efficiency}
%In the gain medium of the proposed system, photons generated on it propagate within the cavity, forming the resonant beam. Then, resonant beams carry the energy and signal arrive at the receiver. 
%We can use the beam power output on the reflector M2 as the figure of merit to evaluate the transmission performance. 
In the gain medium of the proposed optical system, photons are generated and subsequently propagate within the cavity. This process results in the formation of a resonant beam. The resonant beam, functioning as a carrier, transmits both energy and signal to the designated receiver.

\textcolor{blue}{To assess the transmission efficacy of the system, the beam power output at reflector M2 serves as a crucial figure of merit. 
%In the proposed system's gain medium, photons are generated and propagate within the cavity, forming a resonant beam. These resonant beams, carrying energy and signal, reach the receiver. The beam power output at reflector M2 serves as a key figure of merit for evaluating the system's transmission performance.
%Moreover, due to the presence of energy loss, there is a threshold power in the system; the beam output is only available when the input power is greater than the threshold condition. In addition, there is a conversion efficiency ratio between the input and output power of the system, which is determined by the process of energy conversion at each stage. 
Additionally, the system exhibits a characteristic threshold power due to inherent energy losses. The output beam is only generated when the input power surpasses this threshold condition. Furthermore, the relationship between input and output power is governed by a conversion efficiency ratio. This ratio is a function of the energy conversion process occurring at each stage within the system.}
%After the processes of energy-absorbing and stimulated radiation, the resonant beam generates and cyclically oscillate in the SSR. In the receiver, part of the beam will emission from the reflector M2 as a function of the external beam. 
Based on the cyclic power principle and materials characteristic of the gain medium~\cite{koechneR2013solid,kuznetsov1997high}, the external beam power can be defined as:
\begin{equation}\label{Pbeam}
	P_{\mathrm{beam}}=\left(P_\mathrm{in}-P_{\mathrm{th}}\right) \eta_{\mathrm{s}},
\end{equation}
where $P_\mathrm{in}$ denotes the input electric power, $P_\mathrm{th}$ expresses the threshold power, and $\eta_{\mathrm{s}}$ is the slope efficiency. 
%The effectiveness of each conversion level is made up of $\eta_{\mathrm{s}}$, which can be expressed as: 

Firstly, $\eta_{\mathrm{s}}$ is related to the system structure such as the reflectivity of M1, which can be depicted as: 
\begin{equation}
	\eta_{s}=\frac{\eta_{pc}\eta_{pa}\lambda_\mathrm{pump}\ln \left(R_{2}\right)}{\lambda_\mathrm{beam}\ln \left(R_{1} R_{2} V\right)},
\end{equation}
where $R_1$ and $R_2$ denote the reflectivity of the reflectors M1 and M2, $\eta_{pc}$ is the pump conversion efficiency, $\lambda_\mathrm{beam},\lambda_\mathrm{pump}$ represent the wavelength of resonant beams and pump beams, $V=V_cV_d$ expresses overall energy losses which mainly \textcolor{blue}{includes} constant loss $V_c$ (absorption dispersion of reflectors and air, internal loss in the gain) and beam diffraction loss $V_d$. 

Secondly, $P_\mathrm{th}$ \textcolor{blue}{is determined} by the gain medium, which can be expressed as:
\begin{equation}\label{Pth}
	P_{\mathrm{th}}= \frac{A_{s}h c lN_{\mathrm{th}} }{\lambda \eta_{pc}\eta_{pa} \tau},
\end{equation}
where $A_s$ expresses the effective active cross-section, $h$ denotes the Planck constant, $c$ is the speed of light in vacuum, $\lambda$ represents the beam wavelength, $l$ is the length of the gain medium, $N_{\mathrm{th}}$ denotes the threshold carrier density in gain medium, $\eta_{pa}$ expresses the pump absorption efficiency, and $\tau$ is the carrier attenuation time. 

Normally, $\tau$ can be expressed as~\cite{kuznetsov1997high}:
\begin{equation}
	\tau=(\alpha+\beta N_\mathrm{th}+\gamma N_\mathrm{th}^{2})^{-1},
\end{equation}
where $\alpha$, $\beta$ and $\gamma$ express recombination coefficients of the gain medium. $N_{\mathrm{th}}$ is related to the gain factor, which can be described as:
\begin{equation}\label{nth}
	N_\mathrm{th}=N_0\exp(g/g_0),
\end{equation} 
where $g_0$ is gain coefficient under low pump power, $N_0$ denotes the transparency carrier density, and $g$ expresses the saturation gain coefficient. 
%According to Fig.~\ref{sys structure}, \textcolor{blue}{the cavity includes} reflectors M1, M2, TIM, and gain medium. Among them, TIM consists of two lenses with \textcolor{blue}{anti-reflective coating}, whose absorption loss to the incident beam can be considered as 0.
\begin{figure}[t]
	\centering
	\includegraphics[scale=0.42]{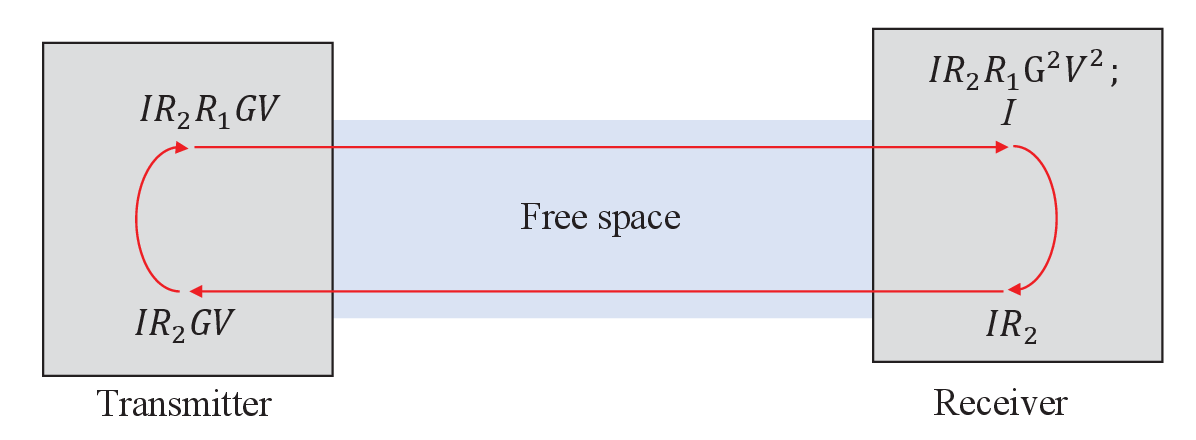}
	\caption{End-to-end energy cycle ($I$: initial energy density; $R_1,R_2$: reflectivity of M1,M2; G: gain factor; $V$: Path loss)}
	\label{Powercycling}
\end{figure}
%Among them, M1, M2 and gain medium will changes the energy while the TIM only changes the phase of the beam.
%To define $g$, we need to analyze the energy saturation condition where the gain and loss are in a balance

To define the parameter $g$, \textcolor{blue}{it is essential to analyze the energy saturation condition, characterized by an equilibrium between gain and loss within the system}~\cite{koechneR2013solid}.
According to Fig.~\ref{sys structure}, \textcolor{blue}{the spatially separated resonator} cavity includes reflectors M1, M2, TIM, and gain medium. %Resonant beams propagate in the cavity and passes through these components.
%In the process of resonant beam propagating within the cavity, optical elements such as M1 will absorb or reflect part of beam causing energy loss. In contrast, energy gains on the gain medium. 
\textcolor{blue}{During the propagation of the resonant beam within the cavity, optical elements, notably M1, either absorb or reflect a portion of the beam, resulting in energy losses. Conversely, energy is gained in the gain medium.}
%Moreover, energy is lost during energy conversion.
%Among them, TIM consists of two lenses with anti-reflective coating.
%The resonant beam propagates in the cavity and passes through these components, obtaining gain in the gain medium and outputting power on M2. During this process, 
%Thus, after one round-trip, if the beam returns to the origin point and its energy is unchanged, we can conclude that the system achieves the balance of loss and gain. This equilibrium condition is called a saturation condition 
Consequently, if the beam, after completing one round-trip within the cavity, returns to its point of origin with its energy unaltered, it can be inferred that the system has attained a state of equilibrium between loss and gain. This specific equilibrium state is referred to as the saturation condition. 
From Fig.~\ref{Powercycling}, after one-trip energy cycle, the initial energy density will become $I R_{2} R_{1} G^{2} V^{2}$ from $I$. When the saturation condition is satisfied, the $I R_{2} R_{1} G^{2} V^{2}$ should equal to $I$, and parameters have the relationship as:%the saturation condition can be expressed as \cite{Hodgson2005Laser,koechneR2013solid}:
\begin{equation}\label{threshold condition}
	R_{1} R_{2} G^2V^2=1,
\end{equation} 
where $G=\exp(\Gamma gl)$ \textcolor{blue}{expresses} the overall gain factor (each time the beam travels through the gain, the beam energy intensity will increase by $G$ times), %$g$ represents the gain coefficient per unit length of the gain medium, 
$R_1$ and $R_2$ can be considered as the beam emission ratio of M1 and M2. 
At this time, $g$ can be depicted as:
\begin{equation}\label{g}
	g = -\frac{\ln(R_1R_2V^2)}{2l}.
\end{equation}
Further, taking the \eqref{g} into \eqref{nth}, the $N_\mathrm{th}$ can be denoted as:
\begin{equation}\label{nth2}
	N_\mathrm{th}=N_0(R_1R_2V^2)^{-(2g_0l\Gamma)^{-1}},
\end{equation} 
where $\Gamma$ is the longitudinal confinement factor. 
%\textcolor{blue}{Normally, the value of $R_1$ is set to 1, and the value of $R_2$ needs to be designed for optimal power output (APPENDIX A).}   
%and $\alpha$ represents the internal loss coefficient per unit length. When the energy accumulates to a certain level, part of the beam will exit as a function of laser.
%When the left side of \eqref{threshold condition} is greater than 1, the radiation of the appropriate frequency will quickly build up until it becomes so large that the exciting transition will consume the high-level energy and reduce the value of $g$. If the gain of each pass just balances the internal and external losses, a steady-state can be achieved.
%After the beam compressed by the TIM, the resonant beam can effectively enter the size-limited gain medium, which makes the micron-level semiconductor gain medium can be applied in the system. 

Finally, based on aforementioned formulas, the  $P_\mathrm{beam}$ presented in \eqref{Pbeam} can be depicted. \textcolor{blue}{At the end,} the  beam transmission efficiency can be defined as:
\begin{equation}\label{eatall}
	\eta_b=\frac{\eta_{pc}\eta_{pa}\lambda_\mathrm{pump}\ln \left(R_{2}\right)}{\lambda_\mathrm{beam}\ln \left(R_{1} R_{2} V\right)} \left(1-\frac{A_{s}h c lN_{\mathrm{th}} }{\lambda \eta_{pc}\eta_{pa} \tau P_\mathrm{in}}\right).
\end{equation}

\subsection{Beam Transmission Description and Diffraction Loss}
%Resonant beams carrying the energy and signal propagate within the SSR. 
To analyze the resonator structure, the process of beam transmission in the resonator should be described at first. In this paper, we adopt beam vectors and transmission matrices to define the beam propagation. 
Specifically, based on the beam \textcolor{blue}{physics} characteristics (straight line propagating), we can depict it as  $\vec{r}=(x,\theta_x)$, where $x$ expresses the position parameter on the coordinate axis and $\theta_x$ represents the propagation direction of the beam. 
%Then, we assume that a beam at the origin is transmitting in free space, from $x_1$ to $x_2$, and its direction angle of transmission is $\theta_1$. Using the beam vector, it can be expressed as $\vec{r_1}=(x_1,\theta_1)$. Then, after the beam travels to $z$ position along the light path, the beam vector becomes $\vec{r_2}=(x_2,\theta_2)$. 
\textcolor{blue}{Assuming a beam originates at a specific point and propagates in free space from \(x_1\) to \(x_2\), with an initial transmission angle denoted as \(\theta_1\), the beam's trajectory can be represented using vector notation. Initially, the beam vector is expressed as \(\vec{r_1} = (x_1, \theta_1)\). As the beam progresses along its path to a position \(z\), its vector representation evolves to \(\vec{r_2} = (x_2, \theta_2)\), indicating the change in both position and transmission angle. }
The process of vectors conversion from $\vec{r_1}$ to $\vec{r_2}$ can utilize matrix $\mathbf{M}$ to express, which is $\vec{r_2}=\mathbf{M}\vec{r_1}$. Moreover, when the beam passes through a series of optical components, the entire process can be expressed as the matrices of these components multiplied in the corresponding order. 

%The retro-reflector has the ability to reflect the incident beam of any direction back to be parallel to the original direction, and is the core element of the system's practice self-alignment function. 
%Utilizing the above beam vectors and transmission matrices, we can define the optical elements and beam transfer in the cavity. 
By employing the aforementioned beam vectors and the transmission matrices, it is possible to define the optical elements and characterize the beam transfer within the cavity. 
Firstly, we adopt it to depict the reflectors. 
The reflectors introduce in the scheme involving in a \textcolor{blue}{convex lens (focal length: $f$)} and a reflective mirror. %The lens is a convex lens whose focal length is $f$. 
The reflective mirror is located at the exit pupil of the lens with distance $d$, and its surface is flat \textcolor{blue}{(infinite focal length). Using the typical matrices' expression of lenses and mirror~\cite{koechneR2013solid},} the reflector can be described as: 
\begin{equation}
		\begin{aligned}
			\mathbf{M}_r=&
			\left[\begin{array}{ll}
				1 & f \\
				0 & 1
			\end{array}\right]
			\left[\begin{array}{cc}
				1 & 0 \\
				-1 / f & 1
			\end{array}\right]
			\left[\begin{array}{ll}
				1 & d \\
				0 & 1
			\end{array}\right]
			\left[\begin{array}{ll}
				1 & 0 \\
				0 & 1
			\end{array}\right]\\
			&\left[\begin{array}{ll}
				1 & d \\
				0 & 1
			\end{array}\right]\left[\begin{array}{cc}
				1 & 0 \\
				-1 / f & 1
			\end{array}\right]\left[\begin{array}{ll}
				1 & f \\
				0 & 1
			\end{array}\right]\\
			=&\left[\begin{array}{cc}
				-1 & 0 \\
				1/f_r & -1
			\end{array}\right],
		\end{aligned}
\end{equation}
where 
\begin{equation}
		f_r=\frac{f^2}{2(d-f)}.
	%f_{\mathrm{r}}=f^2 /\left(\frac{2 l}{f^{2}}-\frac{2}{f}\right)
\end{equation}
Reflectors can reflect incident beams back toward the parallel path of the original one, which is precondition of the self-alignment~\cite{9425612}. 
To realize ideal retro-reflect in the proposed reflectors, we design $d=f$, which denotes the mirror is put at the focal point of the lens. 
%retro-reflectors with $\mathbf{M}_r$ present an ideal telecentric cat’s eye that will 

\begin{figure}[t]
    \centering
    \includegraphics[scale=0.57]{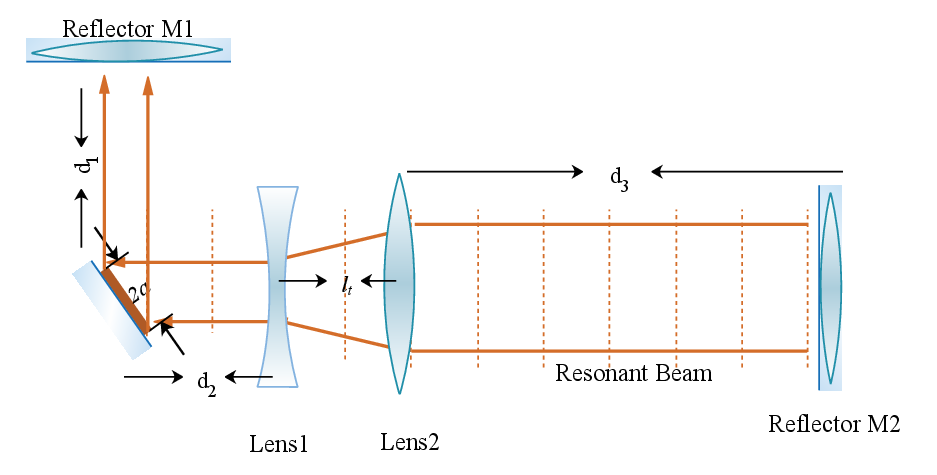}
    \caption{Resonator structure ($d_1,d_2,d_3,l_t$: elements distance; $a$: radius of gain medium)
    }
    \label{PG}
\end{figure}

%The TIM is introduced to compress resonant beams, we apply the matrix to express the TIM. 
\textcolor{blue}{To effectively compress resonant beams, we have introduced the TIM technique and employed matrix representation to characterize and describe the TIM.}
%Subject to material properties, the size of the gain medium is usually relatively small in order to achieve high conversion efficiency. Thus, the beam size is usually greater than the gain medium, which causes unignorable beam loss. 
From Fig.~\ref{PG}, lenses L1, L2 with focal length $f_1$, $f_2$ compose the TIM, which can be depicted as:
\begin{equation}\label{Mt}
	\begin{aligned}
		\mathbf{M}_{\mathrm{TIM}} 
		=&
		\mathbf{M}_\mathrm{L_2}\mathbf{M_{l}}\mathbf{M}_\mathrm{L_1}\\
		=&
		\left[\begin{array}{cc}
			1 & 0 \\
			-\frac{1}{f_{2}} & 1
		\end{array}\right]
		\left[\begin{array}{cc}
			1 & l_t \\
			0 & 1
		\end{array}\right]
		\left[\begin{array}{cc}
			1 & 0 \\
			-\frac{1}{f_{1}} & 1
		\end{array}\right]\\
		=&
		\left[\begin{array}{cc}
			\frac{1}{M} & l_t \\
			0 & M
		\end{array}\right],
	\end{aligned}
\end{equation}
where $M=-f_1/f_2$. %From \eqref{Mt}, the beam 
%If a beam with vector $\vec{r_{1}}=(x_1~0)$ enters the TIM, according to \eqref{Mt}, it will become $\vec{r_{2}}=(x_1/M~0)$ after being modulated by the TIM. From a macro perspective, the beam is toward to light axes with ratio $M$. If $M>1$, it denotes that the beam is compressed. 
%In this paper, we proposed a high-efficiency RB scheme with semiconductor gain medium and TIM. To simplify the analysis process, we use typical curved mirrors instead of retro-reflectors as the reflectors on both sides of the cavity. To achieve reliable and durable SWIPT, the resonant cavity needs to keep stable. At this time, the cavity can restrict the resonant beam overflow and limit the beam to travel back and forth between the transmitter and receiver \cite{Baues1969Huygens,kogelnik1965imaging}.
%System stability represents the ability of the system to operate normally over a long time. 
%When the structure parameters meet the stable condition, it proves 

%To obtain the stability conditions of the cavity, we need to define the propagation process of resonant beam in the cavity at first. 
\textcolor{blue}{Building on the previously mentioned matrices and referring to Fig.~\ref{PG}}, we can define the one trip beams transmission in the resonator. Taking reflector M1 as the starting point, the beam propagating in the cavity will pass through lens 1, lens 2, and reflector M2 in sequence. This process can be expressed using the following matrix representation: %Using matrices to express those process which is:
%Building on the previously mentioned matrices and referring to Fig.~\ref{PG}, we can define the transmission of beams during a single pass through the resonator. Starting at reflector M1, the beam traverses the cavity, sequentially passing through lens 1, lens 2, and then reflector M2. This process can be expressed using the following matrix representation:
%By substitution of the matrix expression depicted in Fig.~\ref{TIM}, the propagation of light in the cavity can be expressed as: 
\begin{equation}\label{transfer matrix1}
\begin{aligned}
\mathbf{M_{c}}= & \mathbf{M_{M2}}\mathbf{M_{d_3}}\mathbf{M_{L_2}}\mathbf{M_{l}}\mathbf{M_{L_1}}\mathbf{M_{d_2}}\mathbf{M_{d_1}}\mathbf{M_{M1}}\\
=&\left[ \begin{array}{cc}
-1 & 0 \\
\frac{1}{f_{r2}} & -1 \\\end{array} \right ]
\left[ \begin{array}{cc}
1 & d_3 \\
0 & 1 \\\end{array} \right ]
\left[ \begin{array}{cc}
1 & 0 \\
-\frac{1}{f_2} & 1 \\\end{array} \right ]
\left[ \begin{array}{cc}
1 & l_t \\
0 & 1 \\\end{array} \right ]\\
&\left[ \begin{array}{cc}
1 & 0 \\
-\frac{1}{f_1} & 1 \\\end{array} \right ]
\left[ \begin{array}{cc}
	1 & d_2 \\
	0 & 1 \\\end{array} \right ]
\left[ \begin{array}{cc}
1 & d_1 \\
0 & 1 \\\end{array} \right ]
\left[ \begin{array}{cc}
-1 & 0 \\
\frac{1}{f_{r1}} & -1 \\\end{array} \right ],
\end{aligned}
\end{equation}
where $\mathbf{M_{M1}}$, $\mathbf{M_{M2}}$, $\mathbf{M_{L1}}$, and $\mathbf{M_{L2}}$ express the matrix of reflector M1, M2 with effective curvature factors $\rho_1$ and $f_{r2}$, and lenses L1, L2 with focal length $f_1$ and $f_2$. 
$\mathbf{M_{d_1}}$, $\mathbf{M_{d_2}}$, $\mathbf{M_{d_3}}$, and $\mathbf{M_{l}}$ express the beam transfer in the space with distances $d_1$, $d_2$, $d_3$ and $l_t$. %Note that we can use typical curved mirrors instead of retro-reflectors as the reflectors on both sides of the cavity to simplify the analysis process. 

%Among them, $M, d_1, f_1$ and $\rho_{2}$ are system structure parameters, and $d_2$ is the environmental variable which expresses the space distance between the transmitter and receiver. Therefore, through the stable cavity inequality, we can get: 1) the variation range of the system's transmission distance $d_2$ at different system structure parameters, and 2) the relationship between the structure parameters when the $d_2$ is determined.
%\begin{figure}[t]
%    \centering
 %   \includegraphics[scale=0.8]{ figure//Powergeneration2.eps}
  %  \caption{Cavity model for the expression of the threshold condition }
   % \label{PG2}
%\end{figure}
%According to section \Rmnum{2}, 
%through the comparison between beam spot size and aperture size, we can judge the degree of the beam divergence.
%beam spot can depict the degree of the beam divergence, and through the comparison between beam spot size and aperture size
%To define the diffraction loss, the distribution of the beam in the vertical propagation direction should be depicted. Beam spot is the distribution of the beam in the vertical propagation direction, and its radius is usually used to evaluate the lateral amplitude of the beam at a certain point.
\textcolor{blue}{To accurately determine the diffraction loss, it is essential to depict the beam's distribution along the vertical propagation direction. 
The beam spot effectively characterizes the beam's vertical distribution, with its radius being a commonly used metric to evaluate the beam's lateral amplitude at a given point. Given that the intensity cross-section of the beam is circular, we utilize the spot radius as the key parameter for this evaluation.}
%The beam spot can characterize this vertical distribution, and its radius is commonly utilized to assess the beam's lateral amplitude at a specific point. Since the intensity cross-section of the beam is circular, we adopt the spot radius to evaluate it.???
%to depict the beam spot size.
%The resonant cavity of the BC-RBC system consists of three elements: M1, M2, and gain module. 
%To compare beam spot size and geometric boundary of the elements
%They all have geometric boundary. 
%We need to analyze the 
%We deduce the expression of the beam spot radius on M1 and M2 at first.
According to \cite{Baues1969Huygens,magni1987multielement}, the beam spots radius on the reflectors of the resonant cavity can be defined by $\mathbf{M_c}$. 
Furthermore, given that the gain medium is strategically positioned on the transmitter side, and considering that the distance $d_1$ is relatively small, it can be inferred that the spot size on the gain medium is approximately equivalent to the spot size on reflector M1.
%Besides, since the gain medium is designed to be located on the transmitter, and $d_1$ is small. It can be considered that the spot size on the gain is approximately equal to the spot size on M1. 
In this case, the beam spot radius on the gain medium can be given by: 
%\begin{equation}\label{mode spot1}
%	\begin{gathered}
%		\textcolor{blue}{\omega_{g} = \sqrt{\frac{\lambda}{\pi}\sqrt{-\frac{B_cD_c}{A_cC_c}}},}
%	\end{gathered}
%\end{equation}
\begin{equation}\label{mode spot1}
\left\{
\begin{aligned}
\mathbf{M_c}=&\left[ \begin{array}{cc}
	A_c & B_c \\
	C_c & D_c \\\end{array} \right ]\\
\omega_{g} =& \left(-\frac{\lambda^2 B_cD_c}{\pi^2 A_cC_c}\right)^{1/4}
\end{aligned}
\right.,
\end{equation}
where $A_c,~B_c,~C_c$, and $D_c$ are the matrix elements of $\mathbf{M_{c}}$ which is depicted in \eqref{transfer matrix1}; $\lambda$ is the wavelength of the resonant beam. %Then, we can utilize $\omega_{g}$ to evaluate the performance of beam compression by TIM. 
Further, the beam diffraction loss $V_d$ can be defined as~\cite{cao2018analysis}:
\begin{equation}
	V_d=1-\exp(-2a^2/w_g^2),
\end{equation}
where $a$ denotes the radius of gain medium.

\subsection{Electric Power output  and Spectral Efficiency}
%According to Section II.A, the external beam with signal will output from the reflector M2 at the receiver. Then, the beam splitter can divide the external beam into two. One part will be transferred to the PV cell for electric energy harvesting (EH). The other part will be caught by the APD for data receiving. 
\textcolor{blue}{As outlined} in Section II.A, the external beam carrying the signal will emerge from reflector M2 at the receiver's end. \textcolor{blue}{Subsequently}, a beam splitter is employed to divide this external beam into two distinct paths. \textcolor{blue}{One portion is directed towards the photovoltaic (PV) cell, facilitating electric energy harvesting (EH). The other portion is captured by the Avalanche Photodiode (APD) for the purpose of data reception.}
%When the beam lasing out from M2, we get external laser beam loaded with information at the receiver. Then, the laser beam will be split, one part is connected to PV to realize electric energy output; the other part is connected to avalanche photodiode (APD) to receive the data.

\emph{1) Electric power output}: 
Firstly, part of the external beam separated by the beam splitter is transmitted to the surface of the photovoltaic cell through a homogenizing waveguide. Then, the beam will be converted into electrical power output through photoelectric conversion. This process can be briefly expressed by the following formula \cite{zhang2018adaptive}:
\begin{equation}\label{PV}
\left\{
\begin{aligned}
&P_{p}=\mu P_\mathrm{beam},\\
&P_{\mathrm{E}_\mathrm{out}}=\eta_{p} P_{p}+P_\mathrm{pth},
\end{aligned}
\right.
\end{equation}
where $\mu$ is the power split ratio, $P_{p}$ expresses the beam power received by the PV. The parameter $\eta_{p}$ is the slope efficiency of photoelectric conversion, and $P_\mathrm{pth}$ is the threshold power of PV. 
%It is worth noting that, to simplify the simulation process, according to the experimental results of \cite{zhang2018distributed}, $P_{\mathrm{E}_\mathrm{out}}$ is expressed by a linear model, where the PV is under ideal heat dissipation and receiving area conditions. 
It should be noted that, in order to simplify the simulation process, $P_{\mathrm{E}_\mathrm{out}}$ is represented using a linear model. This approach is based on the experimental findings presented in \cite{zhang2018distributed}, and assumes ideal conditions for both heat dissipation and the receiving area of the PV (photovoltaic) system.
In practice, the beam energy absorbed by PV is limited, it is expected that the conventional linear EH model is only accurate for the specific scenario when the received powers at the receiver is in a constant power range and the nonlinear EH model has a better applicability~\cite{boshkovska2015practical}. Further, the end-to-end energy conversion efficiency is depicted as: %Thus, it is expected that the conventional linear EH model is only accurate for the specific scenario when the received powers at the receiver is in a constant power range.
\begin{equation}
    \eta_\mathrm{E}=\frac{\eta_{p}P_p+P_\mathrm{pth}}{P_\mathrm{in}}.
\end{equation}

\emph{2) Spectral efficiency}:
%The remaining part of the external beam separated by the beam splitter will act on the APD. The APD converts the optical signal into an electrical signal while receiving the data. This process can be depicted as:
\textcolor{blue}{The residual portion of the external beam, segregated by the beam splitter, is incident upon the APD. The APD functions to convert this optical signal into an electrical signal, concurrently facilitating data reception. This conversion process can be illustrated as follows}:
\begin{equation}\label{d1}
\left\{
\begin{aligned}
&P_D=(1-\mu) P_\mathrm{beam},\\
&I_{\mathrm{D}_\mathrm{out}}=\nu P_\mathrm{D},
\end{aligned}
\right.
\end{equation}
where $\nu$ is the optical-to-electrical conversion responsivity of APD. 

Moreover, photoelectric conversion will produce thermal noise and shot noise when APD is receiving data. Among them, thermal noise can be expressed by the following formula \cite{46}:
\begin{equation}
    n^2_{\rm thermal}=\frac{4kTB_x}{R_L},
\end{equation}
where $k$ is the Boltzmann constant, $T$ is the background temperature, and $R_L$ is the load resistor.
In addition, the formula about the shot noise factor is \cite{46} 
\begin{equation}
    n^2_{\rm shot}=2q(I_{\mathrm{D}_\mathrm{out}}+I_\mathrm{bg})B_x, 
\end{equation}
where $q$ is the electron charge, $B_x$ is the bandwidth, $I_\mathrm{bg}$ is the background current.
At this point, the additive white Gaussian noise (AWGN) of the communication module can be defined:
\begin{equation}
    N^2_\mathrm{M}=n^2_{\rm shot}+n^2_{\rm thermal}.
\end{equation}

Finally, we can obtain spectral efficiency of the system as \cite{lapidoth2009capacity}:
%the signal-to-noise ratio of the system \cite{45}:\begin{equation}\label{d3}SNR=\frac{(P_D_{out})^2}{N^2_{M}}.\end{equation}
%Finally, the spectral efficiency of the RB system can be obtain as \cite{44}
\begin{equation}\label{d2}
\widetilde{C}=\frac{1}{2}\log_2(1+\frac{I_{\mathrm{D}_\mathrm{out}}^2}{2\pi eN^2_\mathrm{M}}),
\end{equation}
where $e$ is the natural constant.
\section{Simulation calculations and results analysis}
%In the section above, we proposed a new RBC system combining with the TOM and the semiconductor chip, which has the capability to enhance the energy conversion efficiency of the system and developed the system's analytical model. In this section, to evaluate the transmission performance of the HE-RBC system, we will compare the original RBC with HE-RBC at first. Then, we will study the impact of the structure parameters on the beam-compression and transmission capabilities, and the achievable transmission performance of the HE-RBC system.
In this section, we will evaluate the performance of the proposed system by analyzing the resonant beam distribution, received beam power, output electric power and channel capacity. 
%Before the calculation, we set some constant parameters of the system whose corresponding values are listed in Table I. 
%Besides, to analyze the performance of the energy transmission and communication capabilities, parameters such as photoelectric conversion factor should be fixed, those parameters and their values are also listed in Table I.

\subsection{Resonant Beam Distribution}
In order to match the semiconductor gain, the proposed RB system introduce a built-in TIM, which can modulate the phase of the beam to achieve the beam compression. To evaluate the compression performance, \textcolor{blue}{we apply the beam spot radius defined in the previous chapter}, analyzing its changes with different structure parameters. 

\emph{1) Parameter Setting}: 
The gain medium is composed of semiconductor material $\mathrm{GaAsP}$, which can \textcolor{blue}{generate} light beam with 980 nm wavelength. It is placed obliquely and is 20 mm away from M1.
Based on the Section II.C, we take the value of $f_{r1}$ as infinite ($d = f$). Then, to constrain the beam to maintain par-axial propagation, the $f_{r2}$ of M2 is set as variable. 
%Moreover, we use the maximum end-to-end transmission distance $d_\mathrm{3,Max}$ to represent the extreme value of $d_3$ when the stable cavity condition is satisfied ($d_\mathrm{2,Max}$ is an ideal value which represents the system's transmission capability, in reality, the transmission distance of the system will be reduced by the energy attenuation such as air absorption). 
The TIM is positioned on the side of the gain medium, and its position parameter is set to $d_2=20$ mm relative to the gain chip. The focal length of the concave lens of the TIM is designed as $f_1=$-5 mm. 
We take the end-to-end distances $d_3$ from 2 m to 10 m, and TIM structure parameter $M$ as the inspection point, respectively. 
Finally, considering the beam spot radius will be different at different distances, we take the maximum value of the spot radius $\omega_\mathrm{g,Max}$ on the gain medium to evaluate. 
%Finally, we can analyze the relationship of the $M$ and $f_{r2}$ on $\omega_\mathrm{g,Max}$.
%Based on~\eqref{transfer matrix1},~\eqref{mode spot1},~\eqref{mode spot3}, Table \ref{t1}, Table \ref{t2}, and the parameters determined above, the relationship curves between $\omega_\mathrm{g,Max}$ and $M$ can be obtained.
%In Part A, through comparison, the BCRB system presents a stable compression capability for incident beam. To obtain the compression limit and influencing factors, we conduct further analysis. Since the beam spot radius will be different at different distances, we use the maximum value of the spot radius $\omega_\mathrm{g,Max}$ on the gain module to analyze. We still set $d = 10, 20, 30, 40$ m as evaluation point. Based on~\eqref{transfer matrix1},~\eqref{mode spot1},~\eqref{mode spot3}, Table \ref{t1}, Table \ref{t2}, and the parameters determined above, the relationship curves between $\omega_\mathrm{g,Max}$ and $M$ can be obtained.
\begin{figure}[t]
	\centering
	\includegraphics[scale=0.6]{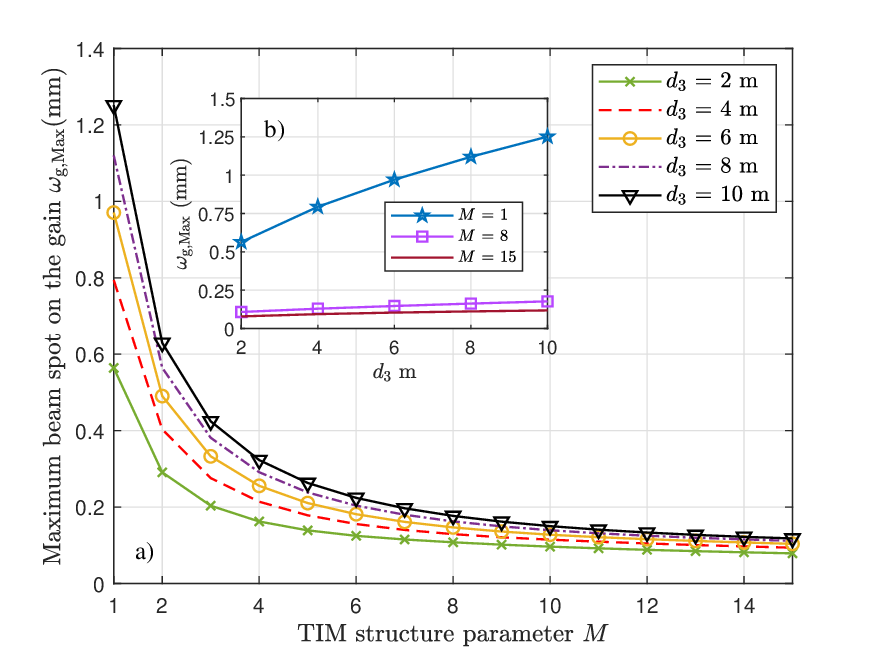}
	\caption{a) Maximum beam radius on the gain versus TIM structure parameter for different curvature of reflector M2; b) Maximum beam radius on the gain versus end-to-end distance }
	\label{d1 vs d2max}
\end{figure}
\begin{figure}[t]
	\centering
	\includegraphics[scale=0.6]{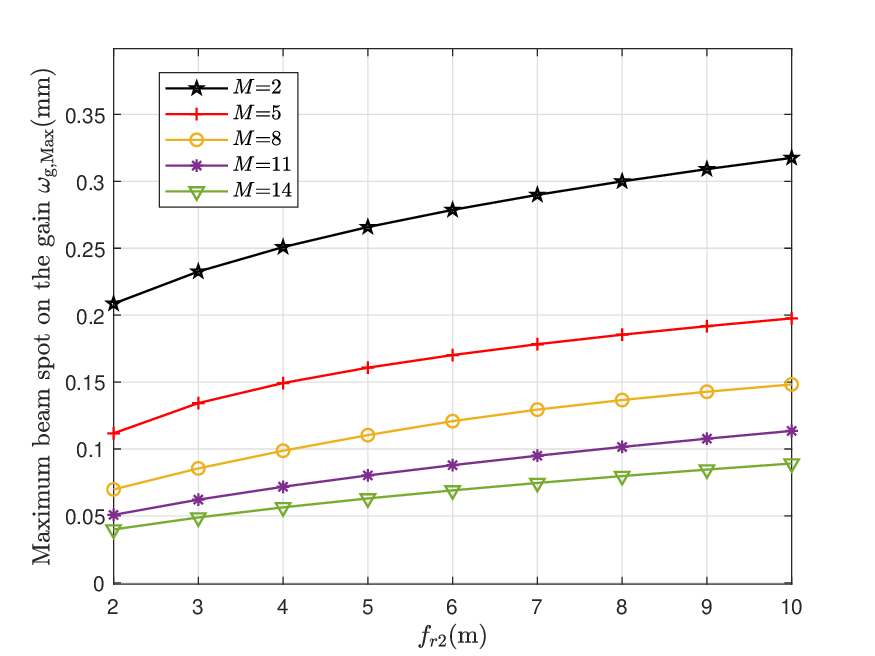}
	\caption{Maximum beam radius on the gain versus $f_r$ parameters of reflector M2 with different TIM structure parameter}
	\label{rho2vsWM}
\end{figure}

\emph{2) Calculation results and analysis}:
Fig.~\ref{d1 vs d2max}(a) \textcolor{blue}{presents the calculation results} which depicts the relationship of the TIM structure parameter $M$ and the maximum beam spot radius $\omega_\mathrm{g,Max}$. As can be seen, when the value of $M$ increases, the $\omega_\mathrm{g,Max}$ shows a downward trend. When $M$ is in the range of 2 to 6, the beam radius decreases rapidly. Then, when $M$ is greater than 10, the decline of the beam spot tends to be gentle, which indicates that compression capacity is limited. Moreover, if the $M$ is beyond the range, the compression tends to be stable, and if $d_3$ takes a large value, the overall $\omega_\mathrm{g,Max}$ will increase. 
In numerical, the \textcolor{blue}{incident} beam can be compressed from 1.2 mm to around 0.1 mm, which is only 1/12 of the original beam. %($M$=1: original scheme without TIM)
Fig.~\ref{d1 vs d2max}(b) \textcolor{blue}{depicts} curves of $\omega_\mathrm{g,Max}$ as a function of $d_3$. As it \textcolor{blue}{presented}, in the original scheme without TIM \textcolor{blue}{(where $M$ equals 1)}, the radius of beam presents a great trend of increase when $d_3$ enhances. In contrast, the proposed scheme with TIM shows a gentle trend of increase. Numerically, the value of $\omega_\mathrm{g,Max}$ in proposed scheme is below 0.25 mm while the original scheme is great then 0.5 mm.
%Above results prove the beam compression ability of TIM, and the compression capacity will increase with the increase of $M$. 
Further, we analyze the influence of $f_r$ parameter on beam compression. Fig.~\ref{rho2vsWM} describes curves of maximum beam radius on the gain versus $f_r$ parameters of reflector M2. 
%As it shown, $\omega_\mathrm{g,Max}$ presents an upward trend with the increase of $f_{r2}$. Then, curves move down as the value of $M$ increases. Overall, the increase of $f_{r2}$ brings a negative effect on beam compression, as the beam spot enlarging. 
\textcolor{blue}{As depicted}, $\omega_{\mathrm{g,Max}}$ exhibits an upward trend as the value of $f_{r2}$ increases. \textcolor{blue}{Subsequently, the curves demonstrate} a downward shift with the increase in the value of $M$. Overall, the increase in $f_{r2}$ tends to have a \textcolor{blue}{detrimental} effect on beam compression, as evidenced by the enlargement of the beam spot.
%the effect brought by $f_{r2}$ is relatively small, specifically when the value of $M$ is large.

From the aforementioned analysis, preliminary conclusions can be obtained that the TIM can effectively compress the beam under different end-to-end distance. Based on Section II.A$\&$B, this ability ensures that the resonant beam can enter the micron-level size semiconductor gain without causing large energy loss. %As a result, the system can achieve long-distance.% and high beam compression at the same time.% by matching the $M$ and $f_{r2}$ reasonably.

\subsection{Received Beam Power}
%\subsection{Energy Output}
According to Section II.A, after the energy accumulating, the external beam will eventually emerge from reflector M2. Then, it will enter the PV and APD respectively under the function of the beam splitter, realizing the electric power output and data reception. Usually, we can adopt the received beam power $P_\mathrm{beam}$ to evaluate the end-to-end beam transmission performance. 
%Firstly, we need to evaluate the power and energy conversion efficiency of the external beam before entering the beam splitter. We set the splitting ratio $\mu$ of the beam splitter to 1, and analyze the total conversion efficiency of the system when it is only used for energy harvesting for evaluating the efficiency improvement of the proposed system compared with the original RB charging system.
%The beam energy refers to the energy of the external beam emitted from the second mirror at the receiver. Because the light beam outside the cavity will split for energy supply and communication. Therefore, the condition of the external beam and the cavity will directly determine the digital energy simultaneous transmission characteristics of the system.
\begin{table}[t]
	\centering
	\caption{Parameters of the gain medium\cite{kuznetsov1997high}}
	\label{t2}
	\begin{tabular}{lll}
		\hline
		\textbf{Parameter} &\textbf{Symbol}&\textbf{Value} \\
		\hline 
		Gain factor&$g_0$&2000 cm$^{-1}$
		\\
		Transparency carrier density&$N_0$&$1.7\times10^{18} \rm cm^{-3}$

		\\
		Longitudinal confinement factor&$\Gamma$&2.0
		\\
Light speed&$c$&3$\times$10$^8$ m/s
		\\
		Planck constant&$h$&$6.6260693\times10^{-34}$J$\cdot$s
		\\
		Monomolecular recombination coefficient&$\alpha$&$10^{7} \rm sec^{-1}$		
		\\
		Bimolecular recombination coefficient&$\beta$&$10^{-10} \rm cm^3/sec$
		\\
		Auger recombination coefficient&$\gamma$&$6\times10^{-30} \rm cm^6/sec$
		\\

		\hline
	\end{tabular}     
\end{table}

\emph{1) Parameter Setting}: The gain medium is made of $\mathrm{GaAsP}$. According to the material properties, we set the characteristic parameters such as light quantum, gain factor, etc. and list them in Table \ref{t2}. Furthermore, we adopt a \textcolor{blue}{typical} laser diode as the pump source whose electro-optical conversion efficiency $\eta_{pc}$ and absorption efficiency $\eta_{pa}$ are 0.6 and 0.85. In addition, the effective active cross-section $A_s$ of the gain \textcolor{blue}{is} set as 3$\times$10$^{-4}$ cm$^2$, the constant loss $V_c$ takes 0.99, the geometric radius of gain medium $a$ is 0.5 mm, the pump beam wavelength is 808 nm, the reflectivity of the mirror in M1 is 0.999, and the end-to-end transmission distance is 10 m. %, and the diameter of the pump area is 100 $\mu$m. 

\begin{figure}[t]
	\centering
	\includegraphics[scale=0.6]{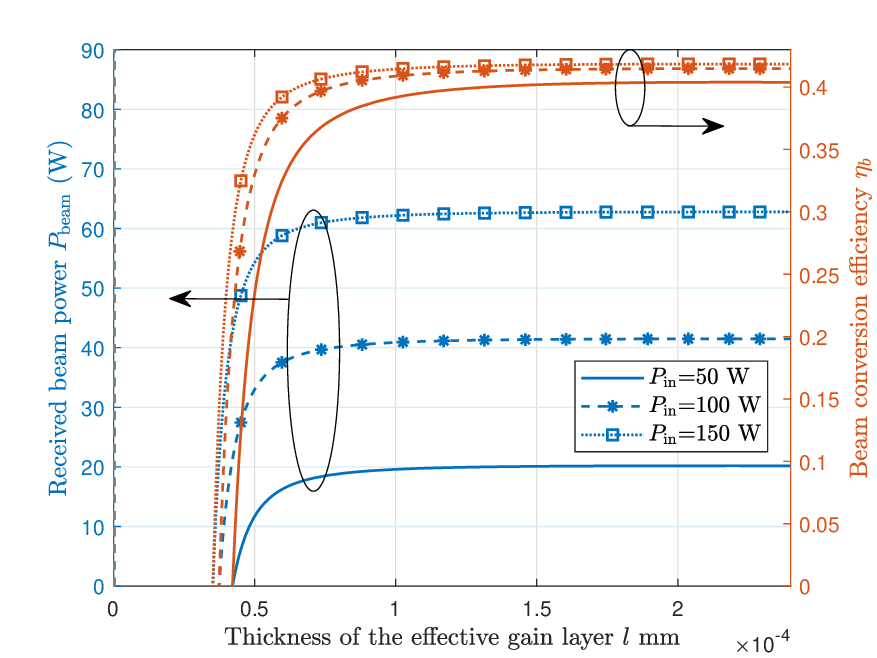}
	\caption{Received beam power and conversion efficiency versus thickness of the effective gain layer}
	\label{ls}
\end{figure}

\emph{2) Calculation results and analysis}:
%Fig.~\ref{ls} describes the relationship of the received beam power, conversion efficiency, and thickness of the effective gain layer. As can be seen, as $l$ increases, $P_\mathrm{beam}$ and $\eta_b$ will also rise. However, the trend of increase tend to be flat when the thickness of the effective gain layer takes a large value. The influence of the $l$ on $P_\mathrm{beam}$ and $\eta_b$ become inconspicuous. 
Fig.~\ref{ls} describes the relationship among the received beam power ($P_\mathrm{beam}$), conversion efficiency ($\eta_b$), and the thickness of the effective gain layer ($l$). It is observed that as $l$ increases, both $P_\mathrm{beam}$ and $\eta_b$ exhibit a rising trend. However, this upward trend tends to plateau when the thickness of the effective gain layer reaches a larger value. Consequently, the influence of $l$ on both $P_\mathrm{beam}$ and $\eta_b$ becomes less pronounced. 
Moreover, when $P_\mathrm{in}$ takes a large value, curves of $P_\mathrm{beam}$ and $\eta_b$ will move up. In numerical, the received beam power can be 60 W and beam conversion efficiency can be 40 $\%$ as $P_\mathrm{in}$=150 W, and the increase of $P_\mathrm{in}$ has great impact on $P_\mathrm{beam}$ while has insignificant impact on $\eta_b$. 
Overall, by \textcolor{blue}{appropriately} increasing the thickness, we can \textcolor{blue}{enhance} the performance of beam transmission.
%Overall, we can improve the performance of beam transmission by raising the thickness properly. 
\begin{figure}[t]
	\centering
	\includegraphics[scale=0.6]{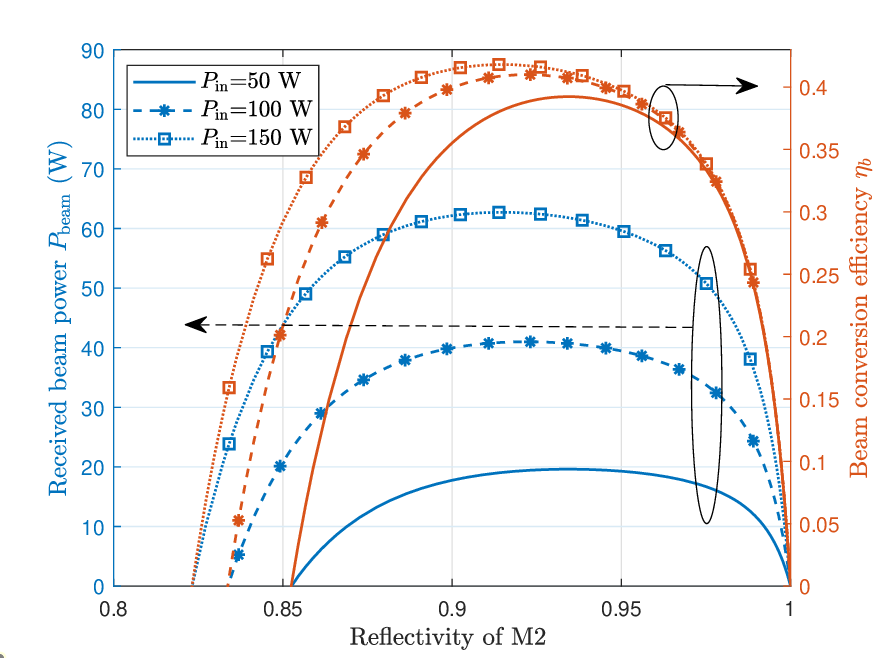}
	\caption{Received beam power and conversion efficiency versus reflectivity of M2}
	\label{R2}
\end{figure}

Fig.~\ref{R2} presents the curves of received beam power and conversion efficiency versus reflectivity of M2. 
%As can be seen from the blue lines, as reflectivity of M2 ($R_2$) changes from 0.8 to 1, both received beam power and beam conversion efficiency will increase and then decrease. Moreover, when input power enhancing, curves of $P_\mathrm{beam}$ and $\eta_{b}$ will advance to the left and ascending. 
\textcolor{blue}{From the blue lines in the graph, it is evident that as the reflectivity of mirror M2 ($R_2$) changes from 0.8 to 1, both the received beam power and the beam conversion efficiency initially increase and then decrease. Moreover, as the input power is increased, the curves representing the beam power ($P_{\mathrm{beam}}$) and the beam conversion efficiency ($\eta_b$) shift to the left and ascend.} 
%Observing the blue lines, %it's evident that as the reflectivity of M2 ($R_2$) varies from 0.8 to 1, both the received beam power and the beam conversion efficiency initially increase and then decrease. Furthermore, with an increase in input power, the curves representing $P_\mathrm{beam}$ and $\eta_{b}$​ shift leftward and upward.
Numerically, the maximum value of $P_\mathrm{beam}$ and $\eta_{b}$ are obtained when the $R_2$ locates at the range from 0.9 to 0.95. 
%The solid red line shows the relationship between $\eta_b$ and $R_2$. The trend of the curve is the same as the power change curve. 
%The trend of the curve is the same as the power change curve. 
%When $R_2$ takes nearly 0.9, the conversion efficiency can reach 0.43. From the curves, we can obtain the influence law of $R_2$ on $P_\mathrm{beam}$ and $\eta_b$, and obtain the optimal value of $R_2$ that satisfies the extreme performance of the system. It is worth noting that when $R_2$ is less than 0.8, there is no power output, which indicates that the loss of the system is greater than the gain at this time, and the energy channel cannot be established.

\begin{figure}[t]
	\centering
	\includegraphics[scale=0.6]{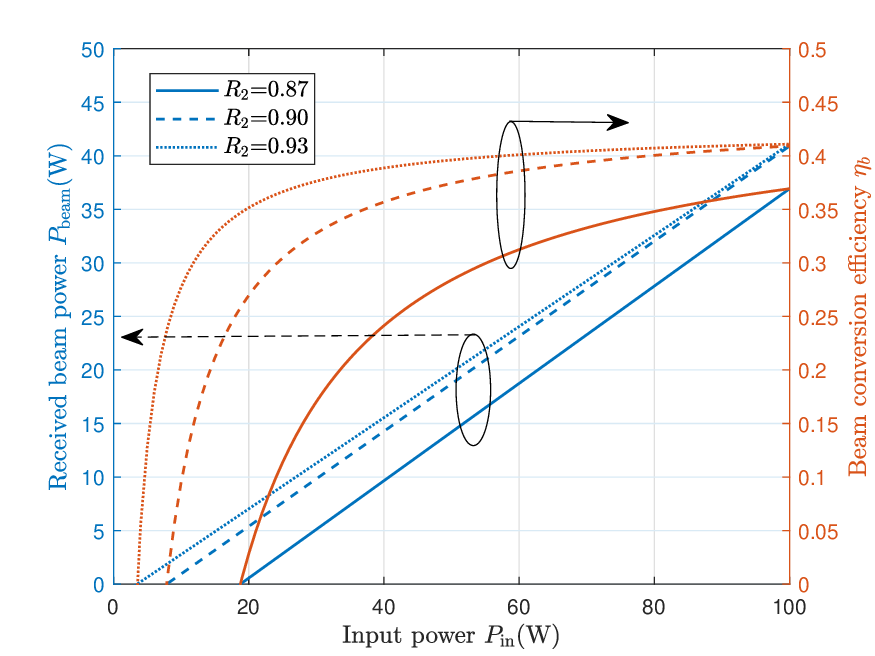}
	\caption{Received beam power and conversion efficiency versus input power}
	\label{Pin}
\end{figure}

\begin{figure}[t]
	\centering
	\includegraphics[scale=0.6]{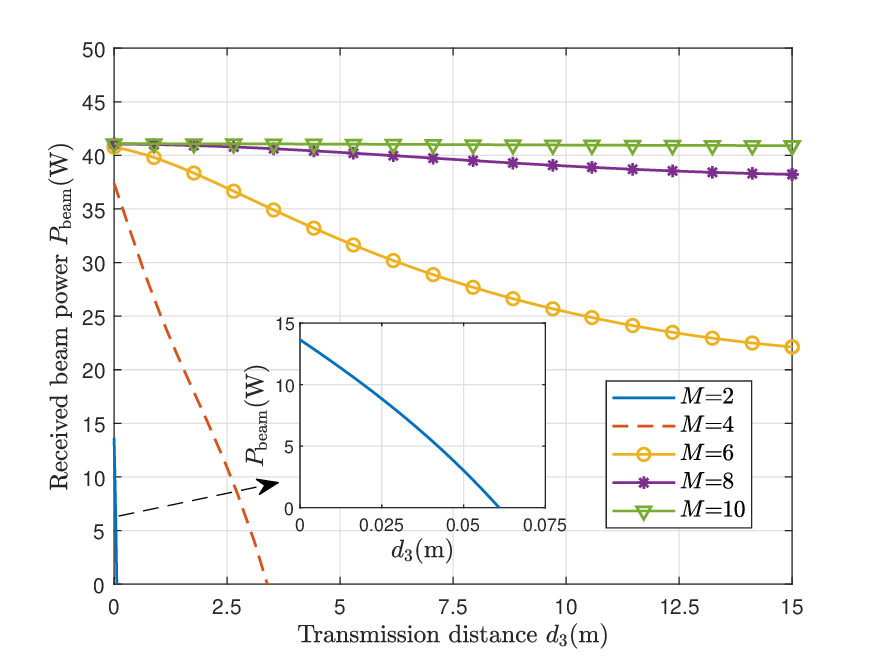}
	\caption{Received beam power versus transmission distance}
	\label{d}
\end{figure}

In Fig.~\ref{Pin}, we analyze $P_\mathrm{beam}$ and $\eta_{b}$ as the function of the input power $P_\mathrm{in}$. According to the blue lines, the $P_\mathrm{beam}$ will increase as $P_\mathrm{in}$ increase, which presents a positive liner relationship with $P_\mathrm{in}$. 
Moreover, the slope and intercept of lines are affected by $R_2$. A large value of $R_2$ is benefit for reducing the threshold power (line intercept). Besides, from the red curves, the $\eta_{b}$ will be also enhanced by $P_\mathrm{in}$ increase. However, different from $P_\mathrm{beam}$, curves are nonlinear and gradually flattens out. 
\textcolor{blue}{
Further, we evaluate the performance of received beam power $P_\mathrm{beam}$ as the function of transmission distance $d_3$, which is shown in Fig.~\ref{d}. As can be seen, curves of $P_\mathrm{beam}$ present a downtrend with the increase of the $d_3$ which proves the negative impact of the transmission loss on beam output (Section II.C). What's more, with the value of $M$ increases, curves become flattened. It indicted that the transmission loss can be restrained by TIM. In numerical, the value of $P_\mathrm{beam}$ can maintain 40W output over 15m distance.
}

%截距

%Fig.~\ref{Powerv1} describes the relationship of received beam power and conversion efficiency versus reflectivity of reflector M2.
%change of the external beam power and the output electrical power on the input power. As can be seen, both $P_\mathrm{beam}$ and $P_\mathrm{out}$ has a linear relationship with the input power. As the input power increases, the value of them will increase. What's more, the threshold power (intercept distance) is small. \textcolor{blue}{For instance, when the value of $R_2$ takes 0.93, the threshold of the system is only a few watts.} Besides, through simulating separately by changing the parameters of $R_2$, we discover that the slope efficiency changes with the $R_2$ changing (Analysis of the impact of $R_2$ on the $P_\mathrm{beam}$ is presented in Appendix~A). When the reflectivity varies from 0.9 to 0.99, the slope efficiency presents an upward trend. \textcolor{blue}{Numerically, when $R_2$ takes 0.93, the value of $P_\mathrm{out}$ can be 13 W at $P_\mathrm{in}$=100~W.}
\subsection{Output Electric Power and Channel Capacity}
%The external beam emitted from M2 will be split by the beam splitter, one part of which enters the PV part for energy output, and the other part enters the APD for signal reading. 
%To evaluate the received beam power, we have analyzed the influence factor of the $P_\mathrm{beam}$ and $\eta_{b}$. Further, the performance of PV and APD for simultaneous wireless information and power transfer need to be analyzed. 
To evaluate the received beam power, we have examined the factors influencing both $P_{\mathrm{beam}}$ and $\eta_b$. Additionally, it is necessary to analyze the performance of the PV cell and the APD in the context of simultaneous wireless information and power transfer (SWIPT).
%after passing the beam splitter, part of the external beam will be caught by the APD for data receiving.
%$\mu$ represents the proportion of light energy entering the PV.

\emph{1) Parameter Setting}: To evaluate the SWIPT performance, the PV and APD structure parameters should be defined. \textcolor{blue}{According} to~\cite{zhang2018distributed}, PV cell structure parameters $\eta_{p}$ and $P_\mathrm{pth}$ are set as 0.3487 and -1.535 W, respectively. 
In the proposed system, the environment temperature is taken as $T$=300 K, the electronic charge $q$ is 1.6$\times10^{-19}$ C, and the Boltzmann constant $k$ is 1.38$\times 10^{-23}$~J/K. The APD for light signal receiving is commercial sensors for 980 nm wavelength. Then, based on \cite{46,8277715,7875119,moreira1997optical}, we set the conversion responsivity of the APD $\nu$=0.6 A/W, the noise bandwidth $B_x$=811.7 MHz, the background current $I_\mathrm{bg}=$5100~$\mu$A, and the load resistance $R_{\mathrm{L}}$=10~K$\Omega$. 
%The structure parameters of APD are defined as $I_\mathrm{bg}=$5100$\mu$A, $\gamma$=0.6 A/W, B=811.7 MHz and $R_{\mathrm{AL}}$=10 K$\Omega$, respectively~\cite{46,8277715,7875119,moreira1997optical}.

\begin{figure}[t]
	\centering
	\includegraphics[scale=0.6]{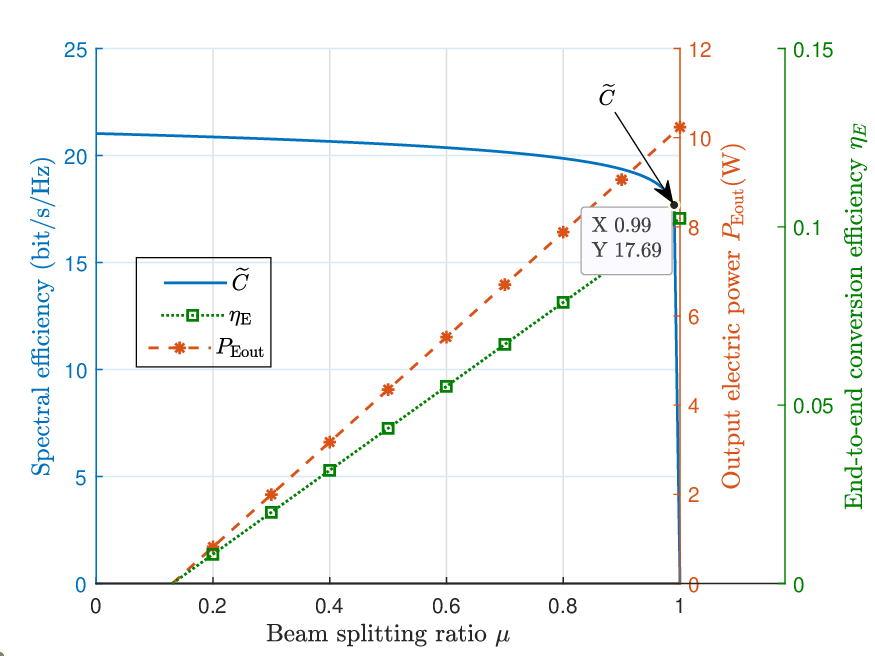}
	\caption{Spectral efficiency, output electric power, and end-to-end conversion efficiency versus beam split ratio ($P_\mathrm{in}$=100 W)}
	\label{compound}
\end{figure}

\emph{2) Calculation results and analysis}:
Fig~\ref{compound} depicts the curves of 
spectral efficiency, output electric power, and end-to-end conversion efficiency versus beam splitting ratio. 
%As can be seen from the blue line, entirety curve presents a downward trend with an increase of $\mu$. The decline of the curve is relatively gentle at the beginning. When $\mu$ is greater than 0.8, $\widetilde{C}$ begins to drop sharply. 
\textcolor{blue}{Observing} the blue line, the entire curve exhibits a downward trend as $\mu$ increases. \textcolor{blue}{Initially}, this decline is relatively gradual. However, when $\mu$ exceeds 0.8, $\widetilde{C}$ starts to decrease sharply.
%Then, from the red and green lines with markers, both $P_\mathrm{Eout}$ and $\eta_{E}$ have positive liner relationship with $\mu$. The value of them will increase with $\mu$ increase. In numerical, the spectral efficiency $\widetilde{C}$ can be 17.69 bit/s/Hz when $\mu$ takes 0.99, where $P_\mathrm{Eout}$ and $\eta_{E}$ can also be high. Thus, we can adopt it to be the beam splitting ratio to achieve high-efficiency SWIPT. 
From the red and green lines with markers, it is evident that both $P_{\mathrm{Eout}}$ and $\eta_{E}$ exhibit a positive linear relationship with $\mu$. Their values increase as $\mu$ increases. 
Numerically, the spectral efficiency $\widetilde{C}$ can reach 17.69 bit/s/Hz when $\mu$ is set to 0.99, \textcolor{blue}{at which point} $P_{\mathrm{Eout}}$ and $\eta_{E}$ are also high. Therefore, this value of $\mu$ can be chosen as the beam splitting ratio to achieve high-efficiency SWIPT.
%What's more, a large input power will make the curve move upward, which causes the overall value of $\widetilde{C}$ to become large. 
%In addition, it can be seen from the curve that the input power will also affect the spectral efficiency. 

\begin{figure}[t]
	\centering
	\includegraphics[scale=0.6]{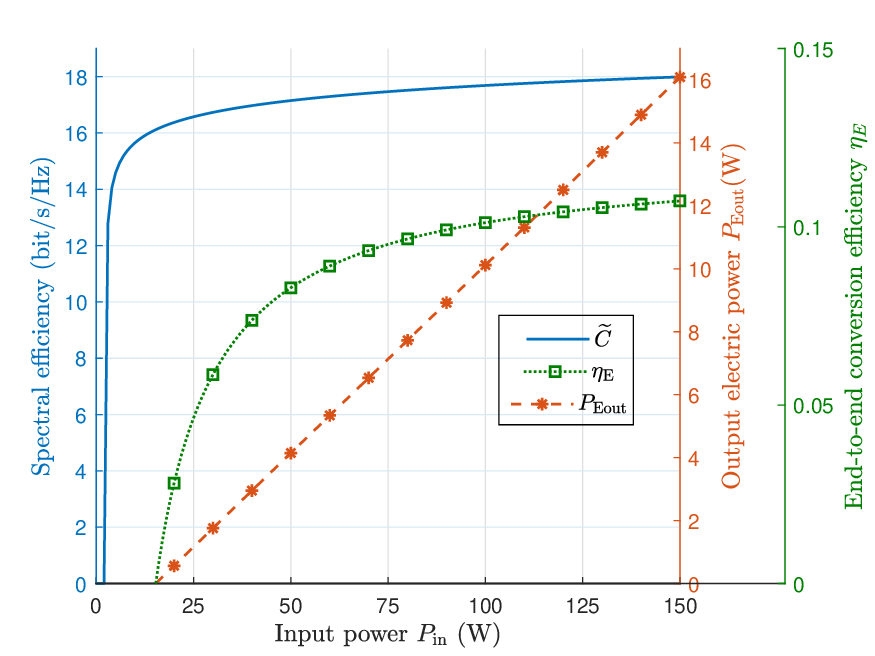}
	\caption{Spectral efficiency, output electric power, and end-to-end conversion efficiency versus input power ($\mu$=0.99)}
	\label{compound2}
\end{figure}

Besides, we also explore the relationship of spectral efficiency, output electric power, and end-to-end conversion efficiency versus input power, which is presented in Fig.~\ref{compound2}. %From the theoretical principle presenting in Section II, 1-$\mu$ represents the proportion of external beam entering the APD, which means the larger $\mu$, the less beam energy is allocated to the APD. Combining the above analysis, we can know that the spectral efficiency of the APD is affected by the amount of beam energy that enters. 
%As can be seen from the blue line, the spectral efficiency will fast rises to a larger value. Overall, the greater the input beam energy, the higher spectral efficiency of the scheme will obtain. However, this influence is relatively limited. With the input power continues to increase, the spectral efficiency of the system will be stable. When $P_\mathrm{in}$ increases from 50W to 150W, the spectral efficiency only increases 1 bit/s/Hz. 
\textcolor{blue}{From} the blue line, \textcolor{blue}{it's apparent that} the spectral efficiency rapidly ascends to a larger value. \textcolor{blue}{In general}, the greater the input beam energy, the higher the spectral efficiency the scheme achieves. However, this effect is somewhat limited. As the input power continues to increase, the system's spectral efficiency tends to stabilize. For instance, when $P_{\mathrm{in}}$ is increased from 50W to 150W, the spectral efficiency experiences only a modest increase of 1 bit/s/Hz.
%From the influence curve of the spectral efficiency, it can be known that the input power and the splitting ratio will affect the spectral efficiency of the system. 
%Taking $\mu=0.01$ (situation that most energy is split to APD) as an instance, 
%Among them, the influence of the splitting ratio is greater, and the influence of the input power is smaller. 
%Furthermore, according to the red and green lines, both the $P_\mathrm{Eout}$ and $\eta_{E}$ have positive relationship with $P_\mathrm{in}$ where the $P_\mathrm{Eout}$ is liner and $\eta_{E}$ is nonlinear. In numerical, $P_\mathrm{Eout}$ and $\eta_{E}$ can up to 16 W and 0.11 at $P_\mathrm{in}$ = 150 W. Generally, through reasonable matching of splitting ratio and input power, the system's data transmission spectral efficiency can reach up to 16-18 bit/s/Hz, and support 0-16 W electric power, which proves that the system has the potential for high-rate and high-power SWIPT. 
Furthermore, \textcolor{blue}{as indicated by} the red and green lines, both $P_{\mathrm{Eout}}$ and $\eta_{E}$ exhibit a positive relationship with $P_{\mathrm{in}}$, where $P_{\mathrm{Eout}}$ shows a linear trend and $\eta_{E}$ demonstrates a nonlinear one. In numerical, $P_{\mathrm{Eout}}$ and $\eta_{E}$ can reach up to 16 W and 0.11, respectively, at $P_{\mathrm{in}} = 150$ W. Generally, by optimally matching the splitting ratio with the input power, the system's data transmission spectral efficiency can achieve 16-18 bit/s/Hz, supporting 0-16 W of electric power. This demonstrates the system's potential for high-rate and high-power SWIPT.
%Generally, the proposed scheme present a remarkable SWIPT performance.

%In addition, the splitting ratio also has a certain impact on the spectrum efficiency. The larger the splitting ratio, the smaller the spectral efficiency of the system.

\begin{table*}[t]
	\renewcommand\arraystretch{1.2} 
	\centering
	\caption{\textcolor{blue}{Comparison of existing SWIPT schemes}}
	\label{t4}
\begin{tabular}{|c|c|c|c|c|c|}
	\hline \textbf{Technology} & \textbf{Ref.} & \textbf{Conversion Efficiency} & \textbf{Spectral Efficiency} & \textbf{Output Power}&\textbf{Transmission Distance} \\
	\hline \multirow{2}{*}{ Visible light } & \cite{ma2019simultaneous} & 0.38$\times 10^{-4} \%$ & 6 $ \mathrm{bit/s} / \mathrm{Hz}$&2.96$\mathrm{~mW}$ & 1.5$ \mathrm{~m}$ \\
	\cline { 2 - 6 } & \cite{abdelhady2020spectral} & 8.44$\times 10^{-5} \%$ & 8 $\mathrm{bit/s} / \mathrm{Hz}$ & 0.38$\mathrm{~mW}$& 3.0$\mathrm{~m}$ \\
	
	\hline \multirow{2}{*}{ Radio frequency } &\cite{lu2014dynamic} & 1.375$\times 10^{-4} \%$ & Not stated & 5.5$\mathrm{~\mu W}$& 15$\mathrm{~m}$ \\
	\cline { 2 - 6 } &  \cite{krikidis2014simultaneous} & 5$\times 10^{-2} \%$ & 7 $\mathrm{bit/s} / \mathrm{Hz}$& 5$\mathrm{~mW}$ &  10$\mathrm{~m}$ \\

	\hline \multirow{2}{*}{ Resonant beam } & \cite{wang2019wireless} & 1$\%$ & Not stated& 2$\mathrm{ W}$ &  2.6$\mathrm{~m}$ \\	
	\cline { 2 - 6 } & This work & 11$\%$ & 18 $\mathrm{bit/s} / \mathrm{Hz}$ & 16$\mathrm{ W}$ & 15$\mathrm{~m}$ \\
	\hline
\end{tabular}
\end{table*}
\subsection{\textcolor{blue}{Summary}}
\textcolor{blue}{
In summary, after numerical evaluation and analysis, we conclude that the proposed RB SWIPT scheme is capable of providing 16 W electrical energy harvesting and 18 bit/s/Hz spectral efficiency for communication, with 11\% energy conversion efficiency and 10 m transmission distance. Table II presents the performance comparison between our system and the existing architectures, clearly demonstrating the advantages of the proposed RB SWIPT in delivering high power, maintaining high spectrum efficiency, and ensuring high energy conversion efficiency over long transmission distances.
%It is clear that the proposed RB SWIPT has benefits in that it can deliver high power while also permitting high spectrum efficiency and energy conversion efficiency over long transmission distance.
}
%%\begin{figure}[t]
%%    \centering
%%    \includegraphics[scale=0.47]{figure/self-al.eps}
%%    \caption{\textcolor{blue}{Self-alignment simulation of the proposed system' cavity by ray tracing, (a) The beam direction is not perpendicular to the TIM surface, and the TIM's lenses are aligned; (b)The beam direction is not perpendicular to the TIM surface, and the TIM's lenses are not aligned.}}
%%    \label{self}
%%\end{figure}
%As can be seen, proposed RB SWIPT has advantages in providing high power, while simultaneously enabling high spectral efficiency and energy conversion efficiency.
%In summary, this numerical evaluation demonstrated that SMIPT is capable of delivering more than 5W charging power and enabling 9:5bps/Hz spectral efficiency for data transfer, with 40-degree FoV under 350W input power and 3m transmission distance. Besides, we reviewed the energy/data transfer and mobility performance of the related SLIPT schemes to highlight the performance with the proposed SMIPT system. Table IV depicts the optimal performance indicators which are satisfied simultaneously of the existing SLIPT systems. SMIPT has advantages in providing high charging power, while simultaneously enabling high spectral efficiency and a large moving range.

\section{Discussion}
\subsection{\textcolor{blue}{Transmission Model by Electromagnetic Field Propagation}} 
\textcolor{blue}{In the aforementioned sections, we have developed end-to-end beam transmission model based on linear optics using beam matrices. 
Additionally, considering the electromagnetic field characteristics of light waves, we can also adopt electromagnetic field propagation to conduct exact analyses for the amplitude and phase distribution of the light field. 
%The description of electromagnetic field propagation is derived by using Maxwell's equations to develop the wave equations for the electric and magnetic fields~\cite{Hodgson2005Laser}. 
The theoretical framework for describing electromagnetic field propagation is derived from Maxwell's equations, leading to the wave equations for electric and magnetic fields, as cited in~\cite{Hodgson2005Laser}. 
Furthermore, according to Huygens-Fresnel principle, during the propagation of light, the field distribution at any point is determined by the coherent superposition of the wavelets of the incident wave at that point~\cite{teperik2009huygens}. 
%Thus, if we know the field distribution of the initial point light wave $u(x,y)$, according to diffraction theory, its field distribution at point s of the transmission path $u'(x,y)$ can be expressed. Then, we can depict the beam transmission in the resonator. Based on it, we can substitute the resonator boundary conditions of the proposed system and describe the light field change of the beam after an end-to-end cycle. Finally, taking the beam generation from the self-reproductive mode~\cite{fox1961resonant} as the judgment condition for stable output, we can obtain the light field distribution at steady state. 
By understanding the field distribution of the initial point light wave, denoted as $u(x,y)$, we can apply diffraction theory to express its field distribution at a specific point 's' on the transmission path as $u'(x,y)$. This approach allows us to accurately represent the beam transmission within the resonator. Building on this, we can incorporate the boundary conditions of the resonator in our proposed system to describe the changes in the light field of the beam over a complete end-to-end cycle. Furthermore, by considering the beam generation from the self-reproductive mode, as mentioned in~\cite{fox1961resonant}, as a criterion for stable output, we can determine the steady-state light field distribution.
%Thus, if the field distribution of the initial point light wave, denoted as $u(x,y)$, is known, its field distribution at a specific point along the transmission path, expressed as $u'(x,y)$, can be determined according to diffraction theory. This enables the depiction of beam transmission within the resonator. Building upon this, we can incorporate the resonator boundary conditions of the proposed system to describe the changes in the light field of the beam after a complete cycle. Finally, by using the beam generation from the self-reproductive mode, as discussed in~\cite{fox1961resonant}, as a criterion for stable output, the steady-state light field distribution can be ascertained. 
\begin{figure}
\centering
\includegraphics[scale=0.74]{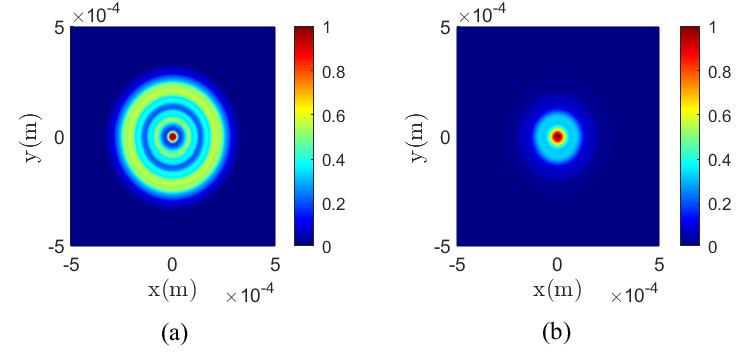}
\caption{Field distribution on the gain medium: (a)without TIM; (b)with TIM }
\label{Edistribution}
\end{figure}
}

\textcolor{blue}{
Fig.~\ref{Edistribution} presents the field distribution on the gain medium based on the electromagnetic field propagation model. As can be seen, the incident beam on the gain medium has been effectively compressed in the proposed system with TIM. In contrast, beam has a large field distribution with low beam quality (uneven distribution) in the original system without TIM. 
%Aforementioned analysis correspond to the results obtained by using the matrix model in section V, which further verifies the feasibility of the proposed scheme. 
This analysis corresponds to the results obtained using the matrix model in Section V, further verifying the feasibility of our proposed scheme.
%Overall, the electromagnetic field propagation can be used to analyze the field distribution of the beam, which is benefit to evaluate the beam's quality, system structure, etc. 
Overall, electromagnetic field propagation analysis is beneficial for evaluating the beam's quality, system structure, etc. 
%However, it is worth noting that in the process of changing the calculation, the fast Fourier transform and the Fox-Li iterative method are adopted in the calculation, but compared with the transmission matrix proposed in the second section of this paper, its computing power requirements and time consumption are still too large. In future research, we will explore in more depth the results and advantages of using this method. 
However, it's important to note that as fast Fourier transform and the Fox-Li iterative method are employed in these calculations, they demand significantly more computing power and time compared to the transmission matrix method introduced in the second section of this paper. Future research will delve deeper into the benefits and results of this method.
%图2展示了基于电磁场传输模型，仿真计算所得的所提出系统增益介质处的光场分布情况，可以看出....
%值得注意的是，在改计算过程中，快速傅里叶变换，和Fox-li迭代法可以辅助计算，但与本文第二节提出了传输矩阵相比，其算力需求和时间消耗仍然过大。在未来的研究中，我们将更深入的探索利用该方法计算的结果和优势。
%According to Section II, When an external excitation is applied to the gain medium, photons are generated and propagated reciprocately in the resonator until the loss and gain are equilibrium. After that, the laser can achieve a stable output. 
%当外部激励作用在增益介质上后，光子会产生，并在谐振腔内循环往复的传播，直至损耗和增益相平衡。之后，激光可以实现稳定的输出。
%根据上述理论，我们就可以分析光在RB系统中的传播，。。。
%
}
%Further, to conduct a more accurate analysis, transmission model by electromagnetic field propagation can be developed. 

%%The structure of the beam splitters are present in Fig.~\ref{beamsplitter}. Cube beam-splitters (Fig.~\ref{beamsplitter} a) are consisted of two classical right angle prisms which can be made by typically glass such as N-BK7. The hypotenuse surface of one prism has beam splitter coating with split ratio $\mu$, and the two prisms are cemented together so that they form a cubic shape. When the incident beam pass through the cube, it will be separate into the reflected beam and the emergent beam. Plate beam-splitters (Fig.~\ref{beamsplitter} b) makes by a glass plate that has been coated on the first surface of the substrate for beam splitter. Plate beam-splitters are set for a 45°, part of the beam is transmitted through the splitter, and the other part of the beam is reflected by the coating and propagates in the vertical incident direction. 
\subsection{System Experiment and Applications}
%According to Fig.~\ref{structure}, several elements are involved in the proposed system. 
\textcolor{blue}{
%In this paper, we propose a scheme for high-efficiency RB charging and communication, provide the analytical model, and evaluate the system performance by simulation. Further, to adopt proposed RB-SWIPT scheme in IoT applications, we will herein discuss the experimental deployment of the system. 
In this paper, we propose a high-efficiency RB charging and communication scheme, provide an analytical model, and evaluate system performance through simulation. For practical implementation of the RB-SWIPT scheme in IoT applications, we should discuss the system deployment and applications in IoT. 
}
%experimental evaluation and applications analysis in IoT will be discussed. 
%we will explore in more depth the results and advantages of using this method. To achieve the system implementation, the effort, cost, and challenges should be discussed. 

\textcolor{blue}{
1) System deployment: 
%As it shows in Fig.~\ref{sys structure} and \ref{PG}, several lenses are involved in the system. Their materials can be N-BK7 glass with high laser damage threshold, and their surface should cover with anti-reflective coating for reducing beam loss. Reflectors M1 and M2 can be designed as cat-eye structure which is depicted in Section II.C. They comprise a convex lens and a flat mirror arranged in parallel, with their pupils overlapping to form a focus-free system. Moreover, reflectors with corner-cube structure (Fig.~\ref{beamsplitter}.(a)) can also be applied in this system for beam retro-reflecting. 
As illustrated in Fig.\ref{sys structure} and Fig.\ref{PG}, the system incorporates several lenses. These lenses can be made of N-BK7 glass, known for its high laser damage threshold, and are coated with an anti-reflective layer to reduce beam loss. Reflectors M1 and M2 are designed with a cat-eye structure, as detailed in Section II.C. This design includes a convex lens and a flat mirror aligned in parallel, with their optical centers overlapping to create a focus-free system. Additionally, reflectors utilizing a corner-cube structure, as shown in Fig.~\ref{beamsplitter}.(a), are also suitable for this system to facilitate beam retro-reflection.
Furthermore, to realize beam separated, optical splitters are introduced in the receiver. Normally, beam splitters such as cube beam-splitters can be used to separate the external beam, which 
\begin{figure}[t]
	\centering
	\includegraphics[scale=0.6]{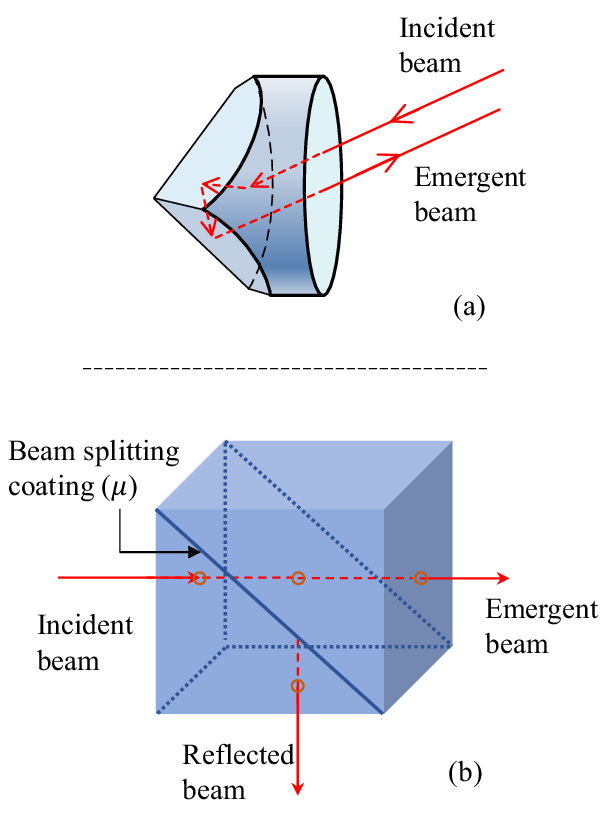}
	\caption{(a) Corner-cube reflector; (b) Cube beam-splitter}
	\label{beamsplitter}
\end{figure}
%The structure of the beam splitters 
are demonstrated in Fig.~\ref{beamsplitter}(b). Cube beam-splitter consists of two classical right angle prisms which can also consist by N-BK7 glasses. The hypotenuse surface of one prism has a coating for beam splitting, and the two prisms are cemented together, forming a cubic shape. %When the incident beam pass through the cube, it will be separate into the reflected beam and the emergent beam, accompanying by the energy assigning under the split ratio $\mu$.  
%Plate beam-splitter (Fig.~\ref{beamsplitter} b) makes by a glass plate that has been coated on the first surface of the substrate for beam splitting. %Plate beam-splitter can be set for 45-degree, 
When beams incident, part of the beam will transmit through it, and the other part of the beam is reflected by the coating and propagates in the vertical incident direction. }

\textcolor{blue}{
The gain module utilized in our system is an open-cavity surface-emitting semiconductor gain, requiring specific customization. As depicted in Fig.~\ref{quantumwell}, the semiconductor gain module is composed of several material regions. Central to these is the Distributed Bragg Reflector (DBR) layer and the gain layer, both of which are fabricated using material growth equipment. This fabrication process involves creating a multi-layer stack structure with alternating high and low refractive indices, along with the quantum well structure previously mentioned. The topmost layer of the module serves as the photon exit for light emission, which can be optimized with an anti-reflection coating to minimize optical losses. The bottom layer functions as a thermal dissipation layer, typically made from a high thermal conductivity metal that is compatible with the semiconductor materials used. 
%In addition, at the receiver, we can use a programmable power supply to change the input state of the pump source by way of amplitude modulation, thus loading the data into the system.
%此外，在接收端我们可以利用可编程电源，通过调幅的方式，改变泵浦源的输入状态，从而将数据加载进系统中。
%Using the above components, combined with the semiconductor gain, 
%根据上述的分析的实验要点，再通过合理的元器件部署，实验平台可以被搭建。
%Based on the analysis of the aforementioned experimental key points, combined with a rational deployment of components, the experimental platform can be constructed. 
%The experimental platform's design, which integrates these key components and considerations, is based on a thorough analysis of these critical experimental factors and a strategic deployment of the various system components.
Drawing on the analysis of the critical experimental factors previously discussed, and through the rational arrangement of components, we can effectively construct the experimental platform.
%PV, and APD depicted in Section II, the proposed RB system for high-efficiency can be built.  
}

\textcolor{blue}{
2) System applications: 
%Based on the analysis presented in Section III, it is evident that the proposed system has the capability to deliver 16W and 18 bits/s/Hz SWIPT over a distance of 15 meters, demonstrating its potential for integration into the IoT network. 
%which prove the great potential of it in the IoT network.%, which can support IoT applications. %We will herein 
%In order to illustrate the practicality of our system, we can provide an example of its application in real-time IoT scenarios.
}
\textcolor{blue}{
%Due to the high mobility, UAVs can serve as air nodes to extend the network coverage and improve the flexibility of IoT topology, which puts forward higher requirements for the endurance and communication capability of the system. 
%The exceptional mobility of unmanned aerial vehicles (UAVs) makes them ideal candidates for serving as air nodes, enabling the extension of network coverage and enhancing the flexibility of the IoT topology. However, this imposes higher requirements on the system's endurance and communication capabilities. To effectively power UAVs and supply dependable data transmission, the RB-SWIPT system can be utilized. 
%To demonstrate the practicality of our system, we offer an example of its application in real-time IoT scenarios. The remarkable mobility of unmanned aerial vehicles (UAVs) makes them exceptionally suited as air nodes, extending network coverage and augmenting the flexibility of the IoT topology. This, however, places greater demands on the system's endurance and communication capabilities. The RB-SWIPT system can be effectively employed to power UAVs and provide reliable data transmission, meeting these enhanced requirements. 
To demonstrate the practical applications of our system, consider its use in real-time Internet of Things (IoT) scenarios, such as with unmanned aerial vehicles (UAVs). UAVs, known for their exceptional mobility, are excellent candidates for acting as aerial nodes. This role not only extends network coverage but also enhances the flexibility of IoT topology. However, employing UAVs in this capacity requires the system to have robust endurance and superior communication capabilities. The RB-SWIPT system is ideally suited for this purpose, as it can effectively power UAVs while ensuring reliable data transmission. 
We introduce a composite UAV system design, comprising a short-endurance UAV, an RB-SWIPT station, and a set of buoy sensors. The UAV's fuselage is equipped with the RB-SWIPT station's receiver module, enabling wireless charging through resonant beam. This feature is crucial when the UAV is tasked with collecting data from sensors, adjusting its flight path, or functioning as a `thing' node in the IoT network. The use of resonant beam technology not only facilitates reliable data transmission but also helps maintain a higher state of charge for the UAVs.
However, the original resonant optical system faced limitations in transmission distance and conversion efficiency due to energy losses, with a maximum transmission distance of only 2.6 meters and a mere 1$\%$ conversion efficiency, as reported in~\cite{wang2019wireless}. These limitations hindered its application in UAVs. Thus, the innovative approach presented in this paper, which integrates a TIM structure and high-efficiency semiconductor gain modules, significantly enhances both the transmission distance and conversion efficiency. This improvement makes the application of the resonant beam system in UAVs feasible. Furthermore, we have developed a viable communication pathway that facilitates simultaneous energy transmission and information delivery. The establishment of a reliable optical communication channel addresses bandwidth and interference challenges, effectively integrating UAVs as functional nodes in IoT applications.
%在原始的共振光系统中，传输距离和转换效率受到能量损失的制约，2.6m的传输距离和1$\%$的转换效率，导致其难以应用于UAV中。本文所提出的方案，基于望远镜结构和高效率的半导体增益模块，在传输距离和转换效率上更有优势，使得共振光系统应用于UAV中成为可能。此外，系统设计了一套可行的通信路径，使得在传能的同时可以进行信息传递。可靠的光通信通道，解决了带宽和干扰的问题，使得UAV可以有效的作为物联网应用节点。
%除上述举的无人机案例外，多种实时物联网终端也可集成RB-SWIPT系统。例如，人体传感器，摄像头，智能门铃等，利用RB-SWIPT系统，通过实时的能量与信号传输，可以有效的解决设备布线，电池占比过大，无线电信号干扰等问题。
%In addition to the UAV case cited above, a variety of real-time IoT terminals can also be integrated with RB-SWIPT systems. For example, human body sensors, cameras, smart doorbells, etc. By implementing the proposed system, issues such as device wiring, excessive battery usage, and radio signal interference can be effectively mitigated through real-time energy and signal transmission. 
%Beyond the UAV application mentioned above, the RB-SWIPT system offers integration capabilities with a wide array of real-time IoT terminals. This includes devices like human body sensors, cameras, smart doorbells, and more. Implementing this proposed system can effectively address common challenges in IoT infrastructure, such as the complexities of device wiring, the problem of excessive battery consumption, and issues with radio signal interference. This is achieved through the system's ability to provide real-time energy and signal transmission, streamlining operations and enhancing overall efficiency. 
Beyond the UAV application mentioned above, the RB-SWIPT system is versatile enough to be integrated with a wide range of real-time IoT terminals. Examples include human body sensors, cameras, smart doorbells, and other IoT devices. The implementation of this system can effectively address common challenges such as the need for extensive device wiring, excessive battery usage, and radio signal interference. This is achieved by enabling real-time energy and signal transmission, thereby enhancing the efficiency and reliability of IoT networks.
%By using proposed system, the problems of device wiring, excessive battery share, and radio signal interference can be effectively solved through real-time energy and signal transmission. 
}
%Several factors may affect the actual performance of the proposed system. First and foremost, the model assumption presented in this paper is based on the beam propagating through the ideal optic components that have the ideal optical performance without aberrations. Thus, when adopting common optic components, there may be deviations between the numerical results and the actual results. To reduce the deviations, special optic components such as aspheric lenses can be adopted.
%Secondly, due to high-density beams are employed as a carrier, the thermal effect should be considered. For example, a large amount of heat accumulation on the gain medium may affect the energy conversion efficiency and produce a thermal lens effect~\cite{chenais2006thermal}. Thus, the efficient heat dissipation such as air cooling, and water cooling can be employed. Moreover, the numerical results in this paper are based on the assumption that the system is located in the clean air. When the system is placed in a complicated external environment, such as a foggy scenario, the beam propagation will be impacted. Thus, the system stability on different environment need to be considered further. 
\section{Conclusions}%\label{conclusions}
%In this paper, we proposed a high-efficiency resonant beam charging and communication system based on the telescope internal modulator and the semiconductor gain medium. Through the transmission matrix, energy cycle, and channel theory, we have established the optical, power, and spectral efficiency models of the system. By systematically analyzing both system structure, energy harvester, and information receiver, we characterize the system’s spot change, energy output, and spectral efficiency. The numerical results illustrate that our system has good stability and can realize long-range digital and energy simultaneous transmission. The system has a better energy output capacity. Compared with the original system, the threshold is lower and the conversion efficiency is increased by 4 times. The system's spectrum efficiency can reach 15bit/s/Hz, which proves the system's communication capability.
%增加讨论的内容
%In this paper, we proposed a high-efficiency resonant beam charging and communication system using the semiconductor gain medium and the telescope internal modulator. Relying on theories of beam transmission, power conversion, energy harvesting, and data receiving, we have established analytical models for the beam propagation, beam power, electric power output, and spectral efficiency of the proposed system. Numerical results illustrate that the system present a remarkable SWIPT performance which can realize 16 W electric power output with 11 $\text{\%}$ end-to-end conversion efficiency, and support 18 bit/s/Hz spectral efficiency for communication.  
In this paper, we have introduced a high-efficiency resonant beam charging and communication system, which utilizes a semiconductor gain medium and a telescope internal modulator. Grounded in the theories of beam transmission, power conversion, energy harvesting, and data reception, we developed comprehensive analytical models. These models address the beam propagation, beam power, electric power output, and spectral efficiency of our proposed system. Numerical analyses demonstrate that our system exhibits exceptional Simultaneous Wireless Information and Power Transfer (SWIPT) performance. It is capable of producing 16 W of electric power output with an 11$\%$ end-to-end conversion efficiency and supporting a spectral efficiency of 18 bit/s/Hz for communication purposes.
%Compared with the original system, it can achieve efficient beam compression, the high power output, energy conversion efficiency enhanced, and high spectral efficiency for high-efficiency resonant beam charging and communication. %Numerical results shows that the proposed RB system can realize 16 W electric power output with 11 $\text{\%}$ end-to-end conversion efficiency, and support 18 bit/s/Hz spectral efficiency for communication. 

\textcolor{blue}{
There are compelling topics that warrant further investigation in future studies.
1) With respect to the performance of the proposed scheme during actual deployment, its influencing factors, and the discrepancies with the simulation results, the experimental analysis of the system can be conducted in the future. 
2) The integration of the proposed system scheme into existing IoT devices, such as system integration, energy reception circuit design, and communication module design, should be thoroughly researched in the future.
}% once the experimental conditions mature
\bibliographystyle{IEEEtran}
\bibliography{references} %----参考文献
\end{document}